\documentclass[11pt]{article}
\usepackage{amsmath}
\usepackage{graphicx}
\usepackage{color}
\usepackage{orcidlink}
\usepackage{hyperref}
\hypersetup{colorlinks=true, linkcolor=blue, citecolor=blue, urlcolor=blue}
\RequirePackage[numbers,sort&compress]{natbib}
\usepackage{subcaption}
\begin{document}
\baselineskip=18pt

\begin{center}
{\LARGE A new black hole coupled with nonlinear electrodynamics  surrounded by  quintessence: Thermodynamics, Geodesics, and Regge-Wheeler Potential }
\end{center}

\vspace{0.3cm}

\begin{center}
{\bf Ahmad Al-Badawi\orcidlink{0000-0002-3127-3453}}\footnote{\bf email: ahmadbadawi@ahu.edu.jo (Corresponding author)}\\ 
{\it Department of Physics, Al-Hussein Bin Talal University, 71111, Ma'an, Jordan}\\
\vspace{0.1cm}
{\bf Faizuddin Ahmed\orcidlink{0000-0003-2196-9622}}\footnote{\bf email: faizuddinahmed15@gmail.com}\\
{\it Department of Physics, University of Science \& Technology Meghalaya, Ri-Bhoi, Meghalaya, 793101, India}\\
\end{center}

\vspace{0.3cm}

\begin{abstract}
We present a new exact solution to the gravitational field equations, where nonlinear electrodynamics (NLED) serves as the matter source in the presence of a quintessence field (QF). This solution describes a static, spherically symmetric black hole in the context of Anti-de Sitter (AdS) space, with the black hole (BH) surrounded by quintessence matter. The resulting black hole solution generalizes the AdS Schwarzschild-Kiselev black hole and reduces to the standard AdS-Schwarzschild black hole in the appropriate limits. Additionally, under certain conditions, this solution recovers the Bardeen-Kiselev regular black hole. We further explore the horizon structure and thermodynamic properties of this new black hole spacetime, accounting for the contributions from both NLED and the QF, including the state parameter of the QF. These contributions modify the spacetime geometry, thereby altering the thermodynamic behavior of the black hole. Moreover, we analyze the geodesic equations and show how NLED and the QF influence the effective potential for massless photon particles. Finally, we study the Regge-Wheeler (RW) potential for this regular black hole and demonstrate how variations in the parameters (NLED and QF) affect the RW potential for fields of different spins: spin-zero ($S=0$), spin-one ($S=1$), and spin-two ($S=2$). To illustrate the effects, we provide graphical representations of the RW potential for both multipole numbers $\ell=0$ and $\ell=1$.
\end{abstract}

{\bf Keywords}: Modified gravity; black holes; Nonlinear Electrodynamics; quintessence field; cosmological constants; Thermodynamics; Geodesics; Regge-Wheeler Potential

\section{Introduction}\label{sec:1}

General relativity is a theory that faces a significant challenge due to the existence of singularities, particularly gravitational singularities that emerge in the context of black holes. These singularities represent regions of spacetime where geodesics break down \cite{KAB}. Consequently, finding ways to avoid singularities in general relativity has become one of the most pressing issues in modern theoretical physics. One promising approach involves regular black holes (BHs), which are solutions to the Einstein field equations that do not exhibit singularities at their cores. Regular black holes are thought to offer an alternative to traditional black holes by eliminating the central spacetime singularity.

A common method for obtaining regular black hole solutions involves coupling general relativity to nonlinear electrodynamics (NLED). One of the key features of NLED is its ability to remove curvature singularities in black hole solutions \cite{ZYF, KAB2}. NLED extends Maxwell's theory to the regime of strong electromagnetic fields, which is especially useful in the study of charged black holes where both gravitational and electromagnetic fields play a crucial role \cite{RGD}. To formulate a consistent theory, it is essential to find a suitable gauge-invariant Lagrangian and its corresponding energy-momentum tensor, which can then be coupled to the Einstein field equations. The first regular black hole solution was derived by Bardeen, who demonstrated that a non-singular geometry could be obtained by modifying the gravitational field with an appropriate energy-momentum tensor that satisfies the weak energy condition \cite{JMB}. Importantly, this solution does not represent a vacuum state; rather, it incorporates matter fields, which is essential for ensuring the regularity of the black hole. Subsequent research showed that the energy-momentum tensor responsible for regular black hole solutions can be interpreted as the gravitational field of a magnetic monopole, originating from a specific form of NLED \cite{ayon,beato,ayon2,ayon3,ayon4}. NLED was proposed in \cite{ni21,ni21a,ni21b} in an attempt to generalize Maxwells theory to strong field regimes. As a result, this theory provides a natural choice for studying charged black holes where we deal with strong electromagnetic and gravitational fields. Later, various regular black hole solutions and their features in spherically symmetric spacetime were proposed \cite{nl1,nl2,nl3,nl4,nl5,nl6,nl7,KAB2}. NLED in the framework of GR allows for the existence of spherically symmetric and static BHs with regular centers \cite{ni20,ni22,ni23,ZYF,nl5}. Recent study has focused on the properties of black holes within the framework of NLED. Furthermore, significant emphasis has been paid to exploring how NLED influences thermodynamic phase transitions and the shadows of black holes \cite{ni26,ni27,ni28,ni29,ni30,ni31,ni32,ni33}. More recent studies have shown that the Kiselev solution, which describes a black hole with a nontrivial energy distribution, can be viewed as an exact solution to the Einstein equations with a power-Maxwell field, either with or without a cosmological constant, depending on the ansatz used-whether for electric charge or magnetic monopole fields \cite{MAD}. 

Numerous authors have been investigated both non-regular and regular black hole solution coupled with nonlinear electrodynamics (NLED) as well as quintessence field, and perfect fluid dark matter and analyzed the results. In Ref. \cite{RKW}, author carried out a comprehensive study of a static and spherically symmetric polymerized black holes metrics. These black holes are motivated by the LQG principles and semi-polymerization technique. In addition, these black holes are not only free from the curvature singularity at the center, geodescially complete and are globally regular, but also free from the blue-shift mass instability as they possess only a single horizon and are globally hyperbolic. These salient features distinguish them from the other regular black holes. Another promising alternative to avoid singularities in black hole solutions is the "black bounce" \cite{AS}, a type of regular black hole characterized by a minimal nonzero area at its core. In this model, singularities are avoided not just by coupling with NLED, but also by introducing an additional form of matter \cite{JCAP}. As the field continues to evolve, numerous other solutions have been proposed, further expanding the possibilities for singularity-free black hole models \cite{BT, CHN}. It is well established that the universe is currently undergoing a period of accelerated expansion, a phenomenon attributed to a mysterious form of repulsive gravity known as dark energy. Several lines of observational evidence support the existence of dark energy. Notably, studies of type Ia supernovae provide compelling evidence for this accelerated expansion \cite{AA1, AA2, AA3}, as do observations of the cosmic microwave background (CMB) \cite{AA4} and large-scale structure (LSS) surveys \cite{AA5, AA6}. However, the precise nature of dark energy remains one of the biggest unsolved problems in cosmology.

Various cosmological models have been proposed to explain dark energy, most of which involve a component of the energy density with negative pressure. The simplest and most widely discussed model is the cosmological constant $\Lambda$, which corresponds to dark energy with an equation of state parameter $w=-1$. Despite its success in fitting observations, the cosmological constant model faces a significant challenge from the perspective of fundamental physics: its observed value is extraordinarily small, a discrepancy known as the "fine-tuning problem" \cite{AA7}. As a result, numerous alternative models have been suggested as candidates for dark energy, many of which involve scalar fields. These include, but are not limited to, quintessence \cite{AA8}, chameleon fields \cite{AA9}, K-essence \cite{AA10}, tachyon fields \cite{AA11}, phantom dark energy \cite{AA12}, and dilaton dark energy \cite{AA13}. The primary distinction between these models lies in the value of $w$, the ratio of pressure to energy density for dark energy. For example, in the case of quintessence, $w$ typically falls within the range $-1 < w <-\frac{1}{3}$. For a comprehensive review of various dark energy models, see \cite{AA14, SF}. Given that dark energy constitutes about $70\%$ of the energy content of the universe, and that black holes are an integral part of our cosmological framework, the study of black holes surrounded by dark energy has become an increasingly important area of research. In this context, this paper focuses on the study of an AdS-Schwarzschild black hole surrounded by quintessence matter. Black holes surrounded by dark energy are thought to play a crucial role in cosmology, influencing both the large-scale structure and the evolution of the universe. Quintessence, as a candidate for dark energy, is typically modeled as a scalar field coupled to gravity, with a potential that decreases as the field value increases \cite{AA8, AA14}. Kiselev \cite{VVK} derived solutions for black holes surrounded by quintessence matter, where the state parameter $w$ lies in the range $-1 < w <-\frac{1}{3}$.

The Hawking radiation of a d-dimensional extension of the Kiselev black hole has been studied by Chen et al. \cite{AA20}, while Ma et al. explored the optical and thermodynamic properties of an Euler-Heisenberg black hole surrounded by perfect fluid dark matter \cite{AA21}. Wu et al. \cite{AA22} investigated a static, spherically symmetric Bardeen-Kiselev black hole with a cosmological constant, which is a solution to the Einstein-nonlinear Maxwell field equations coupled to quintessence. They computed the quasinormal frequencies associated with electromagnetic and gravitational perturbations and analyzed the reflection and transmission coefficients, as well as the absorption cross-section for this black hole. Their findings suggest that the presence of quintessence increases the transmission coefficients. Additionally, in \cite{AA23}, the study of null geodesics and various orbits corresponding to the energy levels of a Schwarzschild-AdS black hole surrounded by quintessence with $w =-\frac{2}{3}$ was undertaken, contributing further insights into the dynamics of black holes in dark energy backgrounds. Numerous studies have investigated various properties of these Kiselev black holes. For example, the quasinormal modes of Schwarzschild black holes surrounded by quintessence matter have been extensively computed \cite{AA15, AA16, AA17, AA18}, and similar studies have been performed for Reissner-Nordström black holes with quintessence matter \cite{AA19}.

Thermodynamics and geodesic properties are two central pillars of black hole physics, both of which have well-established and mature research frameworks. The immense gravitational pull of black holes causes significant bending of light in their vicinity, leading to intriguing optical phenomena. Among these, the shadow of the black hole and the image of its accretion disk are particularly captivating \cite{BB1,BB2,BB3,BB4,BB5}. In the realm of thermodynamics, the study of black holes was significantly advanced by Hawking and Page, who first explored the thermodynamics of black holes in anti-de Sitter (AdS) space-time \cite{BB6}. Since then, numerous important studies have further developed our understanding of black hole thermodynamics \cite{BB7,BB8,BB9,BB10}. Notably, the inclusion of a varying cosmological constant within the extended phase space has led to even more fascinating results, providing deeper insights into the complex behavior of black holes.  

Our current work is inspired by a recent study on Schwarzschild black hole, as presented in \cite{HKS}, but surrounded by nonlinear electrodynamics with the addition of a quintessence field in place of the cloud of strings. In this study, we derive an exact black hole solution in a four-dimensional AdS space-time, where the black hole is coupled to nonlinear electrodynamics as the matter source, under the influence of a quintessence field. The resulting solution smoothly interpolates between the Kiselev-AdS black hole and the standard AdS-Schwarzschild black hole in appropriate limits. We analyze the horizon structure and thermodynamic properties of the black hole, focusing particularly on its stability. These properties are explored both numerically and graphically. The inclusion of nonlinear electrodynamics and the quintessence field modifies the thermodynamics of the black hole, influencing quantities such as temperature, entropy, and heat capacity. Furthermore, we investigate the geodesic motion around this black hole, examining how the nonlinear electrodynamics and quintessence field affect the trajectories of test particles. The impact of these fields on particle motion and stability is thoroughly analyzed. Finally, we investigate the Regge-Wheeler potential for the newly constructed regular black hole with nonlinear electrodynamics and quintessence field. We analyze how the parameters of the black hole influence this potential for fields with spin $0$, $1$, and $2$, respectively.

The paper is organized in a following way: In Sec.~\ref{sec:2}, we obtain an exact regular black hole solution in the presence of a QF using Einstein gravity coupled to NLED in a four-dimensional AdS spacetime and discuss its horizon structure for different parameter values. Sec.~\ref{sec:3} investigates the thermodynamics of the NLED black hole with quintessence. In Sec.~\ref{sec:4}, we investigate the geodesic equation and photon orbit.  The spin-dependent Regge-Wheeler potentials for the regular black hole is presented in Sec.~\ref{sec:5}. Finally, Our findings are briefly summarized in Sec.~\ref{sec:6}.
 
\section{New regular BH solution with NLED under Quintessence}\label{sec:2}

The action describing Einstein's gravity coupled to NLED and surrounded by a QF in 4 dimensions is given by
\begin{equation}
S=\int d^{4}x\sqrt{-\Tilde{g} }\left[ \frac{1}{2\kappa }\left( R-2\Lambda \right) -%
\frac{1}{4\pi }\mathcal{L}^{NE}(F)+\mathcal{L}^{q}\right] 
\end{equation}%
where $\Lambda $ is the cosmological constant, $\Tilde{g}$ is determinant of metric, 
$R$ is Ricci curvature scalar and. $\mathcal{L}^{NE}(F)$ and $\mathcal{L}^{q}$ describe
Lagrangian density for NLED sources and the Lagrangian of the quintessence dark energy respectively. The Lagrangian $\mathcal{L}^{NE}(F)$ is function of $F=\frac{1}{4}%
F_{\mu \nu }F^{\mu \nu }$, where $F_{\mu \nu }=\nabla _{\mu }A_{\nu }-\nabla
_{\nu }A_{\mu }$ is the electromagnetic field strength tensor which is
associated with the gauge potential $A_{\mu }.$

Recall that, the Lagrangian density $\mathcal{L}^{NE}(F)$ satisfies the weak energy
condition as well as spacetime's regularity. Thus, we choose \cite{HKS}
\begin{equation}
\mathcal{L}^{NE}(F)=\frac{Fe^{-\frac{g}{2M}\left( 2g^{2}F\right) ^{1/4}}}{1+\left( 2g^{2}F\right)
^{3/4}}\left( 1+\frac{6\left( 2g^{2}F\right) ^{1/2}}{g\left( 1+\left(
2g^{2}F\right) ^{3/2}\right)}\right),
\label{ne11}
\end{equation}%
where, $g$ and $M$ correspond to the nonlinear charge of a self-gravitating
magnetic field and the mass of the black hole. 

Here, the Lagrangian $\mathcal{L}^{NE}(F)$ satisfies 
\begin{equation}
\lim_{F\rightarrow \infty }\mathcal{L}^{NE}\rightarrow \infty ,\quad
\lim_{F\rightarrow 0}\frac{\partial \mathcal{L}^{NE}}{\partial F}=1.
\end{equation}%
In the weak field limit ($F<<1$), the $\mathcal{L}^{NE}$ refers to Maxwell
electrodynamics. Nonetheless, under the strong field limit, $\mathcal{L}^{NE}$
disappears.

The Lagrangian for the QF is \cite{MER} 
\begin{equation}
\mathcal{L}^{q}=-\frac{6wc}{8\pi }\left( \frac{2F}{g^{2}}\right) ^{\frac{3}{4}\left(
w+1\right) },
\end{equation}%
where $c$ is the normalization constant related to the density of
quintessence $\rho _{q}=-\frac{3cw}{2r^{3w+3}}$ and $w$ is the state parameter with 
$-1 < w < -1/3$.  The stress-energy tensor has
the following components \cite{VVK}
\[
T_{t}^{t}=T_{r}^{r}=\rho _{q},
\]%
\begin{equation}
T_{\theta }^{\theta }=T_{\phi }^{\phi }=-\frac{1}{2}\rho _{q}\left(
1+3w\right) .  \label{qu12}
\end{equation}

The variation of the action with respect to the metric $\Tilde{g}_{\mu \nu }$ gives
the equations of motion for the action, 
\begin{equation}
G_{\mu \nu }=R_{\mu \nu }-\frac{1}{2}g_{\mu \nu }R=T_{\mu \nu }^{NE}+T_{\mu
\nu }^{q},  \label{eom11}
\end{equation}%
\begin{equation}
\nabla _{\mu }\left( \frac{\partial L^{NE}}{\partial F}F^{\mu \nu }\right)
=0,\quad \nabla _{\mu }\left( \ast F^{\mu \nu }\right) =0,
\end{equation}%
where $G_{\mu \nu }$ is the Einstein tensor and $T_{\mu \nu }^{NE}$ and $%
T_{\mu \nu }^{q}$ are the energy-momentum tensor related to the NLED and
QF, respectively. The energy-momentum tensor associated to
NLED is given by%
\begin{equation}
2T_{\mu \nu }^{NE}=\frac{\partial \mathcal{L}^{NE}}{\partial F}F_{\mu \rho }F_{\nu
}^{\rho }-\Tilde{g}_{\mu \nu }\mathcal{L}^{NE}.  \label{emt12}
\end{equation}%
To determine the black hole solution with NLED sources in the presence of
QF, we consider the static spherically symmetric spacetime line
element in 4 dimensions spacetime
\begin{equation}\label{metric}
ds^{2}=-f\left( r\right) dt^{2}+\frac{dr^{2}}{f\left( r\right) }+r^{2}\left(
d\theta ^{2}+\sin ^{2}\theta d\phi ^{2}\right) ,
\end{equation}%
where 
\begin{equation}
f\left( r\right) =1-\frac{2m\left( r\right) }{r}.
\end{equation}%
As a starting point for determining the metric function, we examine
Maxwell's field strength tensor $F_{\mu \nu }$ using the following magnetic
charge choice
\begin{equation}
F_{\mu \nu }=2\delta _{\lbrack \mu }^{\theta }\delta _{\nu ]}^{\phi }Z\left(
r,\theta \right) .
\end{equation}%
Following \cite{ayon2}, we obtain 
\begin{equation}
F_{\mu \nu }=2\delta _{\lbrack \mu }^{\theta }\delta _{\nu ]}^{\phi }R\left(
r\right) \sin \theta .
\end{equation}%
Thus, the non-vanishing component of $F_{\mu \nu }$ is $F_{\theta \phi
}=R\left( r\right) \sin \theta $ and potential $A_{\phi }=-R\left( r\right)
\cos \theta $ \cite{ayon2}. Using $dF=0,$ we obtain $R(r)=g=const.$, where $g$ is the
magnetic charge. Hence, the magnetic field strength is given by 
\begin{equation}
F_{\theta \phi }=g\sin \theta ,\quad F=\frac{g^{2}}{2r^{4}}.  \label{com12}
\end{equation}%
Inserting Eq. (\ref{com12}) into Eq. (\ref{ne11}), then the Lagrangian
density of NLED sources becomes 
\begin{equation}
\mathcal{L}^{NE}(F)=\frac{g^{2}}{2r\left( r^{3}+g^{3}\right) }\left( 1+\frac{6M
}{g}\left( \frac{g^{2}r}{r^{3}+g^{3}}\right) \right)e^{-g^2/2Mr} .
\end{equation}%
Next, we obtain the components of energy momentum tensor given in Eq. (\ref%
{emt12}) as 
\begin{equation}
T_{t}^{NEt}=T_{r}^{NEr}=-2\mathcal{L}^{NE}(F)=-\frac{2M }{\left(
r^{3}+g^{3}\right) }\left( \frac{k}{r}+\frac{3g^3}{r^{3}+g^{3}}%
\right)e^{-k/r} ,
\end{equation} where 
\begin{equation}
  k=\frac{g^{2}}{2M}.  
\end{equation} 
Thus, the $\left( r,r\right) $  components of Eq. (\ref{eom11}) becomes%
\begin{equation}
m^{\prime }\left( r\right) =-\frac{3r^{2}}{2l^{2}}-\frac{Mr}{\left(
r^{3}+g^{3}\right) }\left( k+\frac{3g^{3}r}{r^{3}+g^{3}}\right)e^{-k/r} -\frac{%
3wc }{2 r^{3w+1}}.
\end{equation}%
Integrate the above equation we obtain%
\begin{equation}
m\left( r\right) =-\frac{r^{3}}{2l^{2}}-\frac{Mr^{3}}{r^{3}+g^{3}}e^{-k/r}+%
\frac{c}{ 2 r^{3w}}.
\end{equation}%
Finally,  the metric function for a 4 dimensions black hole with NLED sources in the
presence of QF is
\begin{equation}
f\left(r\right) =1-\frac{2\,M\,r^2}{r^3+g^3}\,e^{-k/r}+\frac{r^2}{\ell^2_{\lambda}}-\frac{c}{r^{3\,w+1}}. \label{m1}
\end{equation}

The derived regular black hole solution (\ref{m1}) is characterized by the mass $M$, cosmological constant $\ell_{\lambda} =\sqrt{-3/\Lambda}$, QF parameters $(c,w)$, magnetic charge $g$, and the deviation parameter $k$.   It is important to note that the presence of the deviation parameter $k$ distinguishes the current black hole metric from the Hayward black hole solution surrounded by quintessence field. Our metric, with the function (\ref{m1}), satisfies several limits as boundary conditions. For instance, setting $g=0$, $k=0$ and $\ell_{\lambda} \to \infty$ results in the Kiselev black hole metric (or Schwarzschild black hole metric with quintessence) \cite{VVK}. Similarly, by choosing $k=0$ and $\ell_{\lambda} \to \infty$, we find the Hayward black hole metric surrounded by quintessence field \cite{OP}. Furthermore, setting $c=0$ and $k=0$ reduces to Hayward-AdS black hole metric \cite{nl4,ZYF2} which is subsequently for $g=0$ results the Schwarzschild-AdS black hole metric \cite{BB6}.

\begin{figure}[ht!]
    \centering
    \includegraphics[width=0.5\linewidth]{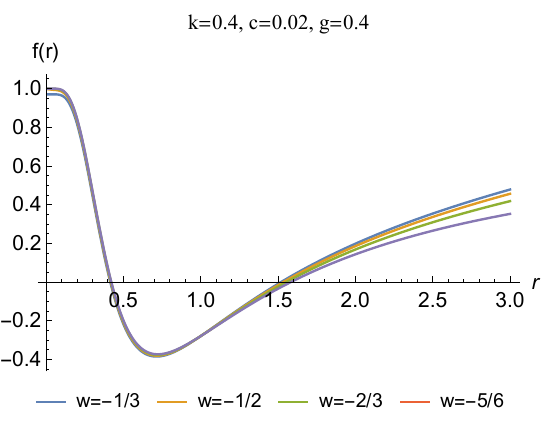}
    \caption{Graphic representation of the metric function (\ref{m1}) with respect to the radial coordinate with different values of $w$. Here $M=1$ and $\ell_{\lambda}=10$.}
    \label{figa1}
\end{figure}

\begin{figure}[ht!]
\begin{center}
\includegraphics[scale=0.9]{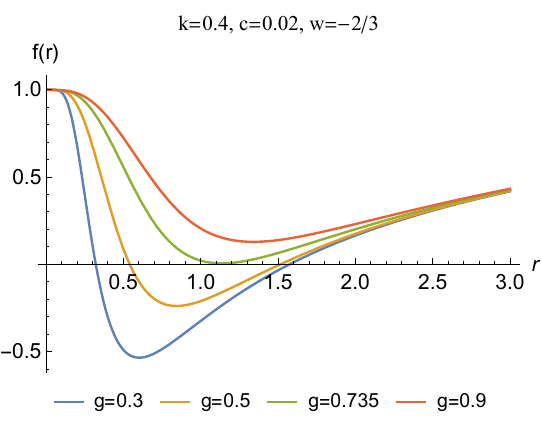}\quad
\includegraphics[scale=0.9]{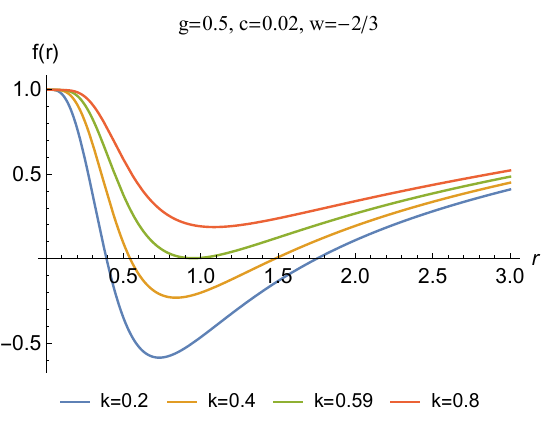}
\hfill\\
\includegraphics[scale=0.95]{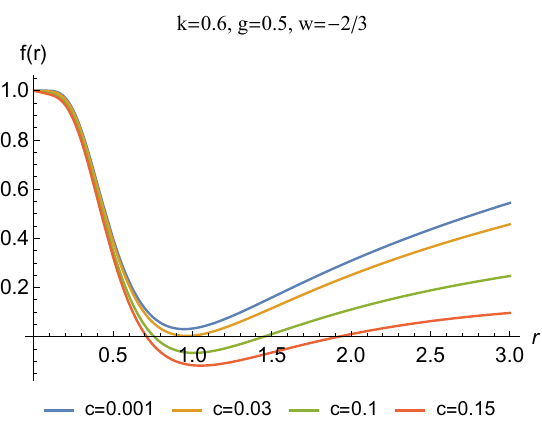}
\end{center}
\caption{Graphic representation of the metric function (\ref{m2}) with respect to the radial coordinate with different values of $g$, $k$ and $c$. Here $M=1$ and $\ell_{\lambda}=10$.}\label{figa2}
\end{figure}

To study how $f(r)$ varies with different parameters, we plot figures \ref{figa1} and \ref{figa2}. Figure \ref{figa1}  shows the metric function with respect to the radial coordinate for various values of the state parameter $w$ by setting the other parameters constant. The state parameter has a relatively minor effect on the metric function. As a result, we will focus on the value $w=-2/3$ in this section. Consequently the metric function becomes 
\begin{equation}
f\left(r\right) =1-\frac{2\,M\,r^2}{r^3+g^3}\,e^{-k/r}+\frac{r^2}{\ell^2_{\lambda}}-cr. \label{m2}
\end{equation} 

Figure \ref{figa2} shows that raising both the deviation parameter $(k)$ and the magnetic charge parameter $(g)$ increases the metric function of the black hole, whereas increasing the normalization constant $c$ decreases it. It is worth noting that the quintessence  parameter behaves in the opposite way as the deviation and magnetic charge parameters do.

The nature of the black hole horizon for the solution obtained in (\ref{m2}) may now be determined using equations $f(r) = 0$. We can get the horizon of the black hole solution and the degenerate horizon. A transcendental function is present in equation (\ref{m2}). As a result, there is no analytic solution, and the problem can only be solved numerically. 

\begin{center}
\begin{tabular}{|c| c c c c| c c c c|}
 \hline 
  &\multicolumn{4}{|c|}{$c=0.01$, $g=0.4$ }&\multicolumn{4}{|c|}{$c=0.02$, $g=0.4$ }
\\ \hline 
$k$ &$r_{-}$ &$r_{+}$ & $\delta$& $r_{ph}$ &$r_{-}$ &$r_{+}$ & $\delta$& $r_{ph}$  \\ \hline
$0.2$ & $0.297$ & $1.738$ & $1.441$ &$2.741$ & $0.296$ & $1.773$ & $1.477$ &$2.786$\\ 
$0.3$ & $0.355$ & $1.621$ & $1.266$ & $2.581$& $0.354$ & $1.654$ & $1.3$ &$2.624$\\ 
$0.4$ & $0.423$ & $1.488$ & $1.065$ & $2.406$& $0.422$ & $1.521$ & $1.099$ & $2.447$ \\ 
$0.5$ & $0.512$ & $1.331$ & $0.819$ &$2.209$& $0.509$ & $1.363$ & $0.854$ & $2.248$ \\ \hline
$g$ &\multicolumn{4}{|c|}{ $k=0.4$ }&\multicolumn{4}{|c|}{ $k=0.4$ }  \\ \hline
$0.2$ & $0.240$ & $1.522$ & $1.282$ &$2.432$& $0.239$ & $1.553$ & $1.314$ &$2.472$\\ 
$0.3$ & $0.323$ & $1.511$ & $1.188$ & $2.423$ & $0.322$ & $1.542$ & $1.22$&$2.464$ \\ 
$0.4$& $0.423$ & $1.488$ & $1.065$ &$2.406$ & $0.422$ & $1.521$ & $1.099$ &$2.447$\\ 
$0.5$ & $0.546$ & $1.447$ & $0.901$ & $2.375$ & $0.544$ & $1.481$ & $0.937$ & $2.417$
\\ 
 \hline
\end{tabular}
\captionof{table}{Numerical results for the  Cauchy horizon
$r_-$, event horizon $r_+$, their deviation $\delta=r_+-r_- $ and the photon sphere for the NLED black hole with QF.} \label{taba1}
\end{center}

In Table \ref{taba1}, it is shown that parameters $c, g,$ and $k$ control the structure of the black hole's horizon as well as the photon sphere  in the absence of cosmological constant, $\ell_{\lambda} \to \infty$. There is no horizon radius for a value of deviation parameters $k > k_c=0.65$ or magnetic charge $g > g_c=0.725$, often known as the critical value. For parameters $k$ or $g$ smaller than their critical value, both a Cauchy horizon $r_-$ and an event horizon $r_+$ exist. Moreover, the horizon and photon sphere of the NLED black hole with QF decrease with increasing $g$ and $k$ but increase with increasing $c$.

\section{Thermodynamics of NLED BH with QF}\label{sec:3}

\begin{figure}[ht!]
\begin{center}
\includegraphics[scale=0.9]{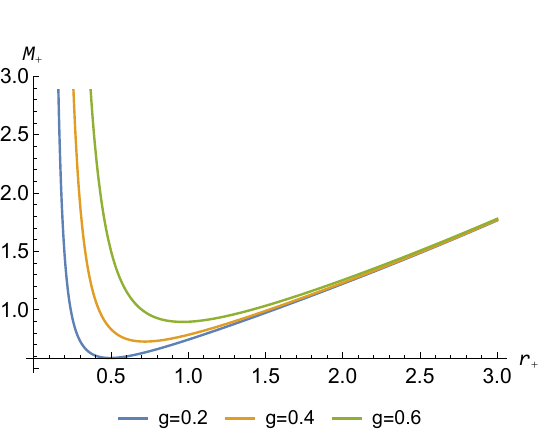}\quad
\includegraphics[scale=0.9]{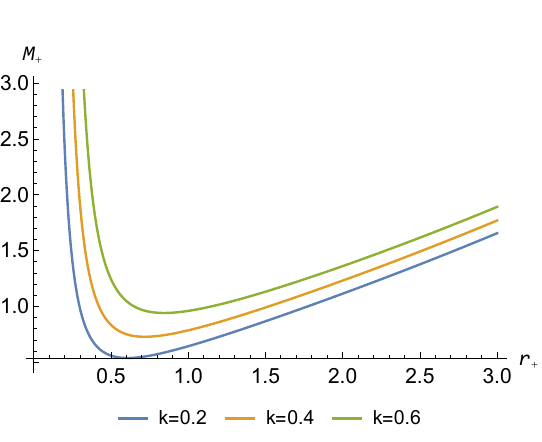}
\hfill\\
\includegraphics[scale=0.95]{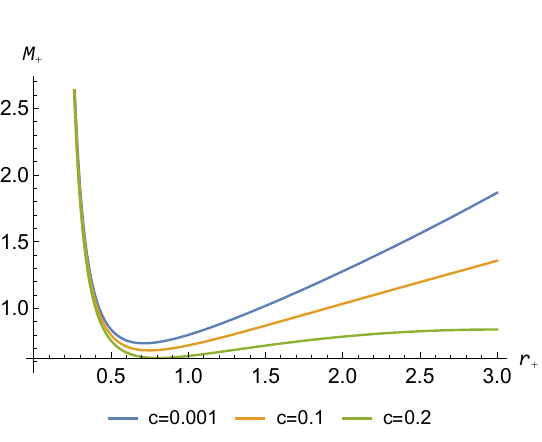}
\end{center}
\caption{The impact of the NLED and QF parameters on the mass of NLED BH with quintessence.}\label{figa3}
\end{figure}

Now we look at the thermodynamic quantities associated with the black hole
solution (\ref{m2}), which is defined by the parameters $M,k,g$ and $c$. In this
part, we discuss the thermodynamic parameters of mass, temperature, entropy,
and specific heat capacity. To calculate the mass of a black hole, put the
metric function $f(r)=0$ at the horizon radius $r_{+}$. This gives%
\begin{equation}
M=\frac{\left(\ell^2_{\lambda}-c\,\ell^2_{\lambda}\,r_{+}+r_{+}^{2}\right) \left(
r_{+}^{3}+g^{3}\right) }{2\,\ell^2_{\lambda}\,r_{+}^{2}}\,e^{-k/r_{+}}.  \label{mass1}
\end{equation}%
Next, we examine the mass function in light of NLED and quintessential
matter. As shown in figure \ref{figa3}, the  NLED and QF  affect the mass
function but does not change its characteristic properties. Its mass reaches a minimum value after decreasing
exponentially and approaching zero at a specific critical horizon radius. In contrast to
the Schwarzschild black hole, the black hole's mass increases linearly as
its radius increases. Furthermore, we see that when $g$ or $k$ increases, so does the mass of the
black hole. Simultaneously, it decreases as the quintessence parameter $c$
increases. 

Next, the temperature associated with a black hole, referred to as the Hawking
temperature, can be calculated by 
\begin{equation}
T_{+}=\left. \frac{1}{4\pi }f^{\prime }\left( r\right) \right\vert
_{r=r_{+}}.
\end{equation}%
Corresponding to black hole solution  (\ref{m2}), the Hawking temperature
is calculated by%
\begin{equation}
T_{+}=\frac{1}{4\pi}\left[\frac{2\,r_{+}}{\ell^2_{\lambda}}+\frac{\left\{\left(k-r_{+}\right)\,r_{+}^{3}+g^{3}\,\left( k+2\,r_{+}\right) \right\}\,\left\{
r_{+}^{2}+\ell^2_{\lambda}\left( 1-c\,r_{+}\right) \right\}}{\ell^2_{\lambda}\,r_{+}^{2}\,\left(r_{+}^{3}+g^{3}\right) }-c\right].  \label{temp1}
\end{equation}

\begin{figure}[ht!]
\begin{center}
\includegraphics[scale=0.9]{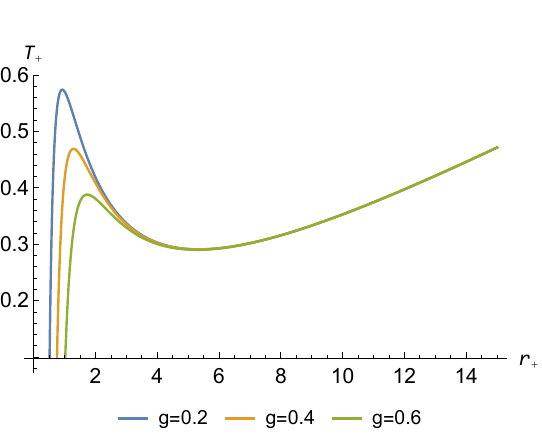}\quad
\includegraphics[scale=0.9]{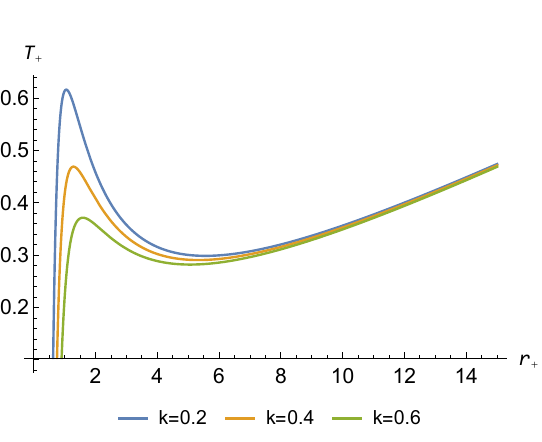}
\hfill\\
\includegraphics[scale=0.95]{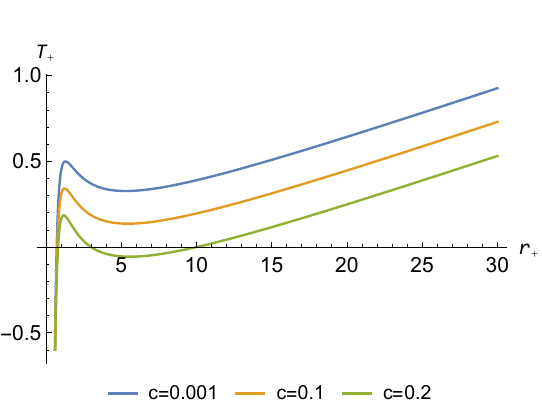}
\end{center}
\caption{The impact of the NLED and QF parameters on the temperature of NLED BH with quintessence.}\label{figa4}
\end{figure}

The Hawking temperature, $T_{+}$, is determined by the black hole parameters  $k,g,$ and $c$, as well as the radius of curvature $\ell_{\lambda}$. In the limit, all three parameters tend to zero ($k=0=g=c$), it reduces to the temperature of the Schwarzschild-AdS black hole as, 
\begin{equation}
T_{+}=\frac{1}{4\pi }\left( \frac{3\,r_{+}}{\ell^2_{\lambda}}+\frac{1}{r_{+}}\right) .
\end{equation}%
We then demonstrate the plot of Hawking temperature versus horizon in figure \ref{figa4}
and examine the effect of the NLED and QF parameters. In all
scenarios, we see that the temperature of this regular black hole climbs
abruptly to reach a maximum value for a specific horizon radius, then
declines exponentially with an increase in horizon radius, and ultimately
increases as the horizon radius approaches infinity. We must emphasize that
as the parameters $k,g,$ and $c$ decrease, so does the black hole's maximum
temperature.

Now, let's look at the other most important thermodynamic number, the black
hole's entropy, using the first rule of thermodynamics. In general, the
explicit form of the entropy $S_+$ can be deduced from the first-law of
thermodynamics  
\begin{equation}
dM=T_{+}dS_{+}.  \label{entr1}
\end{equation}%
This leads to 
\begin{equation}
S_{+}=\int \frac{1}{T_{+}}\frac{\partial M}{\partial r_{+}}dr_{+}.
\label{entr2}
\end{equation}%
Inserting the values of Mass  (\ref{mass1}) and temperature (\ref{temp1})
into (\ref{entr2}), we obtain the entropy of our black hole as%
\begin{equation}
S_{+}=\frac{\pi }{4\,k}\left[ \left( \left( k+r_{+}\right)
\,k\,r_{+}-2\,g^{3}\right)\, e^{-k/r_{+}}-k^{3}\,E_{i}\,\left( \frac{k}{r_{+}}\right) %
\right] ,  \label{entr3}
\end{equation}%
where $E_{i}$ is characterized by exponential integral. This entropy is independent of the QF parameters. Furthermore, it does not comply with the area-law. However, in the absence of $k$ and $g$, this formula perfectly corresponds to the entropy of a Schwarzschild black hole. We may now cross-check the expression of Hawking temperature based on the first law of thermodynamics (\ref{entr1}). 

Now, we have 
\begin{equation}
T_{H}= \frac{\partial M}{\partial S_{+} }=\frac{1}{4\,\pi }\left( \frac{2\,r_{+}}{\ell^2_{\lambda}}+\frac{\left\{\left(
k-r_{+}\right)\, r_{+}^{3}+g^{3}\,\left( k+2\,r_{+}\right) \right\} \left\{
r_{+}^{2}+\ell^2_{\lambda}\,\left( 1-c\,r_{+}\right) \right\}}{\ell^2_{\lambda}\,r_{+}^{2}\left(
r_{+}^{3}+g^{3}\right) }-c\right). \label{entr33}
\end{equation}
This expression is consistent with that obtained by the area-law or tunneling approach in (\ref{temp1}). As a result, our solution for this black hole is consistent with the basic law of thermodynamics.

From the above expression Eq. (\ref{entr33}), we see that the Hawking temperature is influenced by the deviation parameter $k$, the Hayward-like parameter $g$, the cosmological constant $\Lambda=-3/\ell^2_\lambda$, and the constant $c$ for a particular state parameter $w=-2/3$. This result of the Hawking temperature can be used to find the same physical quantity for some well-known black holes under special cases. For example, if we set the deviation parameter $k=0$ and $\ell_\lambda \to \infty$, one will find the Hawking temperature for Hayward black hole metric with quintessence field of state parameter $w=-2/3$. Similarly, setting $k=0=c$, the result reduces for Hayward-AdS black hole. Moreover, if we set $k=0=g$ and $\ell_\lambda \to \infty$, one will find the result of the Hawking temperature for Kiselev black hole with state parameter $w=-2/3$. Finally, setting $k=0=g$ and $c=0$, this result reduces for Schwarzschild-AdS black hole.

To analyze the thermal stability of a black hole as a thermodynamic system, we will focus on the parameters in the NLED and QF. This will be accomplished by using the heat capacity. As a result of investigating the heat capacity, we may determine if the thermodynamic system is locally stable or unstable. In fact, this may be accomplished by examining the sign of the heat capacity, which corresponds to a stable thermodynamic system, as a positive sign of the heat capacity appears, whereas an unstable state is associated with a negative sign. Using the relationship  $C_{+}=\frac{dM}{dT_+}$, it is possible to derive the specific heat capacity of the black hole: 

{\begin{tiny}
\begin{equation}
C_{+}=\frac{2\,\pi\,\left\{e^{-k/r_{+}}\,\left( r_{+}^{3}+g^{3}\right)^{2}\left\{
-r_{+}^{2}\left\{\left( k-3\,r_{+}\right)\, r_{+}^{3}+g^{3}\,k\right\}+\ell^2_{\lambda}\,\left\{
r_{+}^{3}\left\{r_{+}-2\,c\,r_{+}^{2}+\left( c\,r_{+}-1\right)\,k \right.
+g^{3}\,\left\{r_{+}\left( c\,r_{+}-2\right) +k\,\left( c\,r_{+}-1\right) \right\}
\right\}\right\}\right\}}{\left[6\,r_{+}^{7}\left( r_{+}^{3}+4\,g^{3}\right)
-2\,\ell^2_{\lambda}\,r_{+}\left\{g^{6}\left( c\, k\, r_{+}-2\, k-2\, r_{+}\right)+r_{+}^{6}\left(
c\,k\,r_{+}-2\,k+r_{+}\right)+g^{3}\,r_{+}^{3}\left\{ 9\,c\,r_{+}^{2}-4\,k+2\left(
c\,k-5\right) \right\}\right\}\right]}.\label{heatc1}
\end{equation}
\end{tiny}
}

From the above expression (\ref{heatc1}), we see that the specific heat capacity is influenced by the deviation parameter $k$, the Hayward-like parameter $g$, the cosmological constant $\Lambda=-3/\ell^2_\lambda$, and the constant $c$ for a particular parameter of the state $w=-2/3$. This result can be used to find the same physical quantity for some well-known black holes under special cases. For example, if we set the deviation parameter $k=0$ and $\ell_\lambda \to \infty$, one will find specific heat for Hayward black hole with quintessence field of state parameter $w=-2/3$. Similarly, setting the deviation parameter $k=0$ and $c=0$, we find the result for Hayward-AdS black hole. Moreover, if we set the deviation parameter $k=0$, the Hayward-like parameter $g=0$, and $\ell_\lambda \to \infty$, one will find the specific heat for Kiselev black hole metric with state parameter $w=-2/3$. Finally, setting $k=0=g$ and $c=0$, this result reduces for Schwarzschild-AdS black hole. We present one such result below and others are in the same way. Setting the Hayward-like parameter $g=0$ and the deviation parameter $k=0$, we obtain form Eq. (\ref{heatc1})
\begin{equation}\label{heatc2}
C_{+}=\frac{2\,\pi\,r_{+}^{2}\left[3\,r_{+}^{2}+\ell^2_{\lambda}\left(1-2\,c\,r_{+} \right) \right]}{\left(3\,r_{+}^{2}-\ell^2_{\lambda}\right)}.
\end{equation}
Equation (\ref{heatc2}) is the specific heat capacity for Schwarzschild-AdS black hole surrounded by a quintessence field of state parameter $w=-2/3$.

\begin{center}
\begin{figure}[ht!]
\subfloat[$k=0.4, c=0.02$] {\centering{}\includegraphics[scale=0.5]{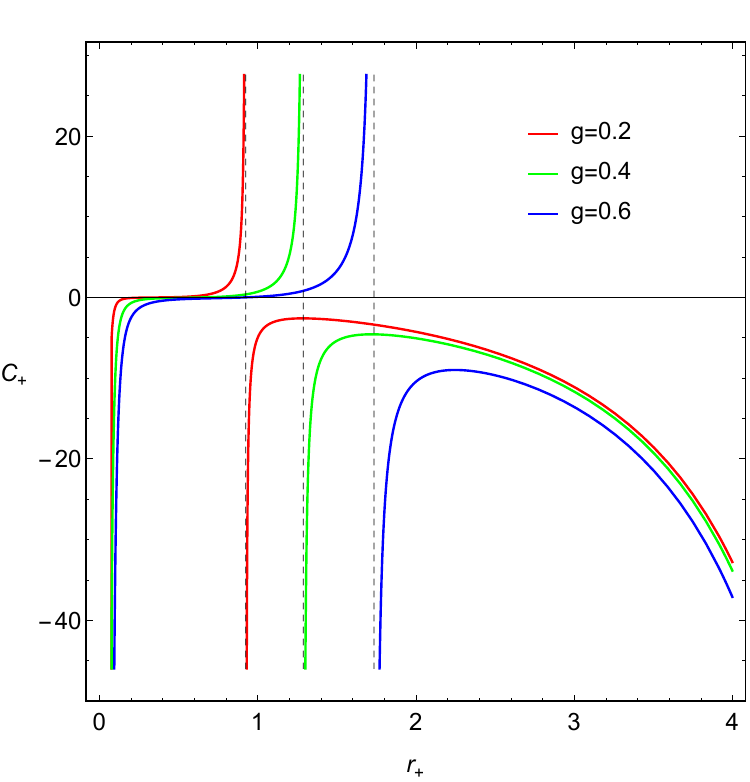}}\quad\quad
\subfloat[$k=0.4,  c=0.02$]{\centering{}\includegraphics[scale=0.5]{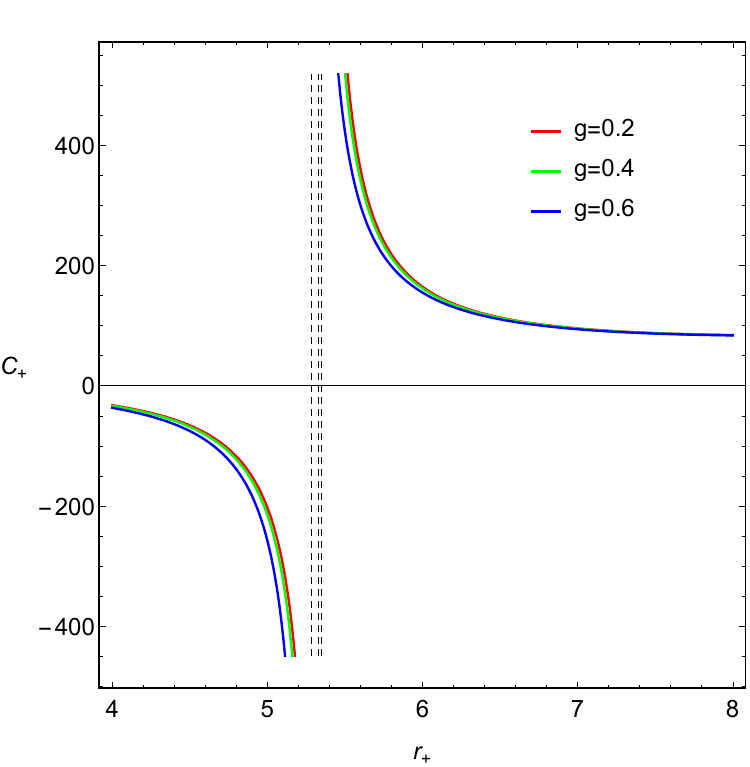}}
\hfill\\
\subfloat[$g=0.4,  c=0.02$]{\centering{}\includegraphics[scale=0.5]{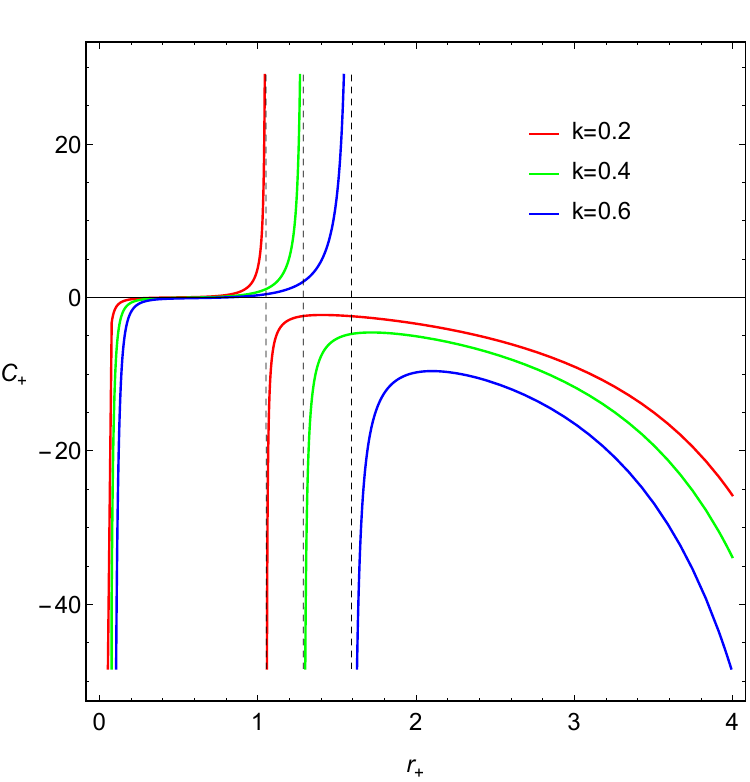}}\quad\quad
\subfloat[$g=0.4,  c=0.02$]{\centering{}\includegraphics[scale=0.5]{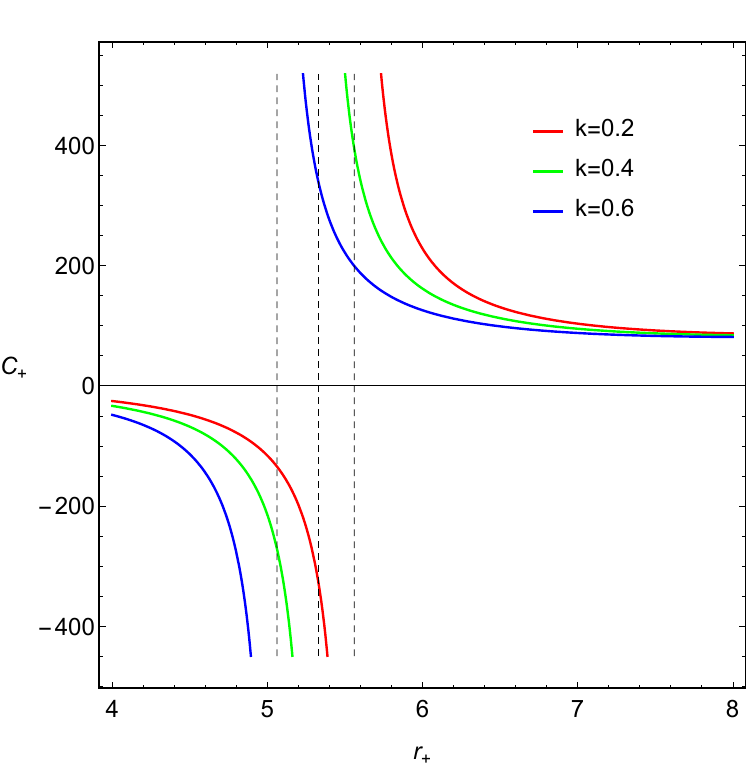}}
\hfill\\
\subfloat[$k=0.4, g=0.4$]{\centering{}\includegraphics[scale=0.5]{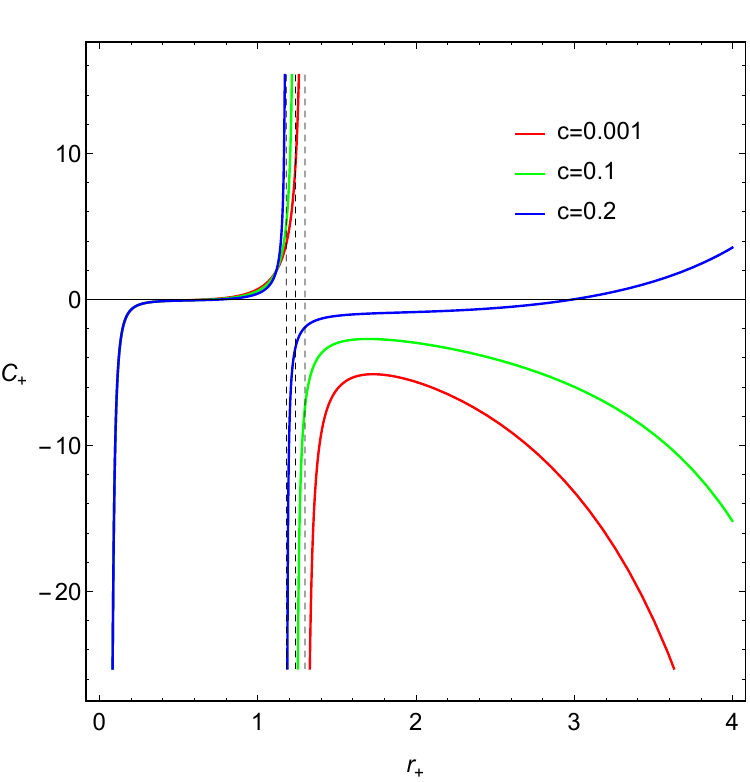}}\quad\quad
\subfloat[$k=0.4,  g=0.4$]{\centering{}\includegraphics[scale=0.5]{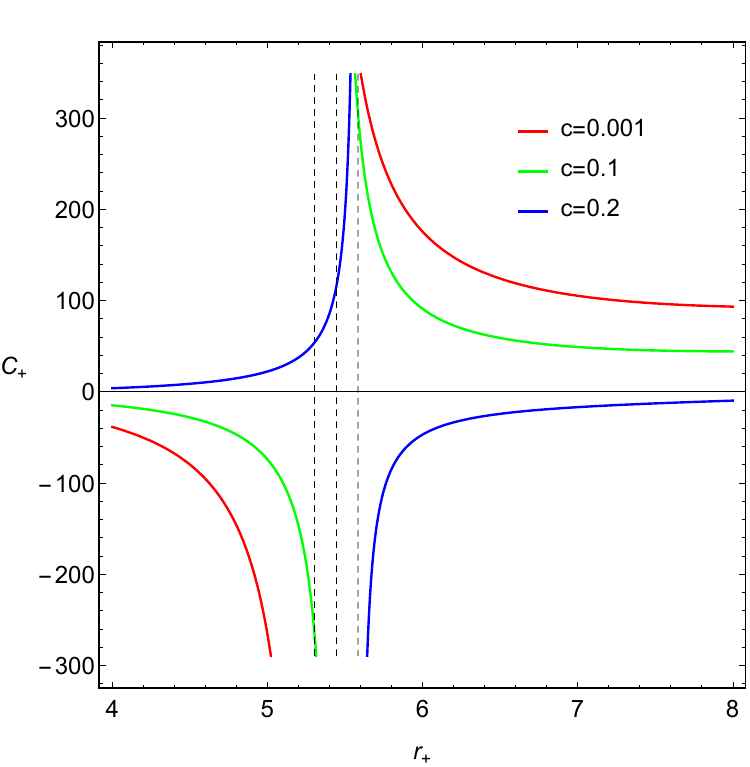}}
\centering{}\caption{Variation of the specific heat capacity $ C_+$ as a function of horizon radius $r_+$ for various values of the parameter space.
 Here we set $M=1, \ell_{\lambda}=10$.}\label{figa5}
\end{figure}
\par\end{center}

A better understanding of heat capacity behavior can be gained by examining figure \ref{figa5}, which shows the variation in heat capacity $ C_+$ as a function of horizon radius $r_+$ depending on various parameter values. The figure depicts the emergence of two divergent points and two physical boundary points that characterize the heat capacity phase transition, which occurs, say at $r_{+1}$ (left panels)   and $r_{+2}$ (right panels). According to the figure \ref{figa5}, we can determine if the black hole system is thermally stable or unstable. For regions $< r_{+1}$ and $> r_{+2}$, the specific heat has a positive value  indicating that the black hole is thermally stable. The region between  $r_{+1}$ and $r_{+2}$ is thermally unstable due to negative specific heat. We note that, these discontinuity points correspond to the turning points of the   Hawking temperature shown in figure \ref{figa4}. 

\section{Geodesic Equations} \label{sec:4}

In this section, we study the geodesic motions of test particles around a new regular black hole with NLED surrounded by QF obtained in section \ref{sec:2}. The metric we formulated in the current paper is given by the following line-element
\begin{equation}
    ds^2=-f(r)\,dt^2+\frac{dr^2}{f(r)}+r^2\,d\Omega^2,\label{qq1}
\end{equation}
where $f(r)$ is given in Eq. (\ref{m1}) as ($\ell^2_{\lambda}=-3/\Lambda$),
\begin{equation}
    f(r)=1-\frac{2\,M\,r^2}{r^3+g^3}\,e^{-k/r}-\frac{\Lambda}{3}\,r^2-\frac{c}{r^{3\,w+1}}.\nonumber
\end{equation}

The Lagrangian function using the metric (\ref{qq1}) at the equatorial plane defined by $\theta=\pi/2$ is given by
\begin{equation}
    \mathcal{L}=\frac{1}{2}\,\Bigg[-f(r)\,\dot{t}^2+\frac{1}{f(r)}\,\dot{r}^2+r^2\,\dot{\phi}^2\Bigg],\label{cc1}
\end{equation}
where dot represents ordinary derivative w. r. t. affine parameter $\tau$ and the function $f(r)$ is given in Eq. (\ref{m1}). One can see from the space-time (\ref{qq1}) that the metric tensor $g_{\mu\nu}$ is independent of both the temporal $(t)$ and azimuthal ($\phi$) coordinates, while depends on $(r, \theta)$. Therefore, there are two Killing vectors $\xi_{t} \equiv \partial_{t}$ and $\xi_{\phi}\equiv \partial_{\phi}$ and hence, two constants of motions are there. 

These two constants of motion represented by $\mathrm{E}$ and $\mathrm{L}$ are given by
\begin{eqnarray}
    &&-\mathrm{E}=-f(r)\,\dot{t}\Rightarrow \dot{t}=\frac{\mathrm{E}}{f(r)},\nonumber\\
    &&\mathrm{L}=r^2\,\dot{\phi}\Rightarrow \dot{\phi}=\frac{\mathrm{L}}{r^2},\label{cc2}
\end{eqnarray}
where $\mathrm{E}$ represents the particles energy and $\mathrm{L}$ is the conserved angular momentum. 

With these, geodesic motion for $r$ coordinate becomes
\begin{equation}
    \dot{r}^2+V_\text{eff} (r)=E^2,\label{cc3}
\end{equation}
where the effective potential is given by
\begin{equation}
    V_\text{eff} (r)=\left(-\epsilon+\frac{\mathrm{L}^2}{r^2}\right)\,\left(1-\frac{2\,M\,r^2}{r^3+g^3}\,e^{-k/r}-\frac{\Lambda\,r^2}{3}-\frac{c}{r^{3\,w+1}}\right),\quad \epsilon=0\quad \mbox{or}\quad -1.\label{cc4}
\end{equation}

From the above expression (\ref{cc4}), it is evident that the effective potential for both null and time-like geodesics is influenced by several parameters. These include the deviation parameter $k$, the Hayward-like parameter $g$, the cosmological constant $\Lambda$, the state parameter $w$ including the constant $c$, and the conserved angular momentum $\mathrm{L}$. 

We have generated Figure \ref{fig:5} illustrating the behavior of the effective potential of null geodesics for different values of various parameters $(g,k,L,\Lambda,w)$. Similarly, in Figure \ref{fig:6}, the effective potential of time-like geodesics for different values of the aforementioned parameters is depicted. 

\begin{center}
\begin{figure}[ht!]
\subfloat[$k=0.1,L=1,\Lambda=-0.1,w=-2/3$]{\centering{}\includegraphics[scale=0.4]{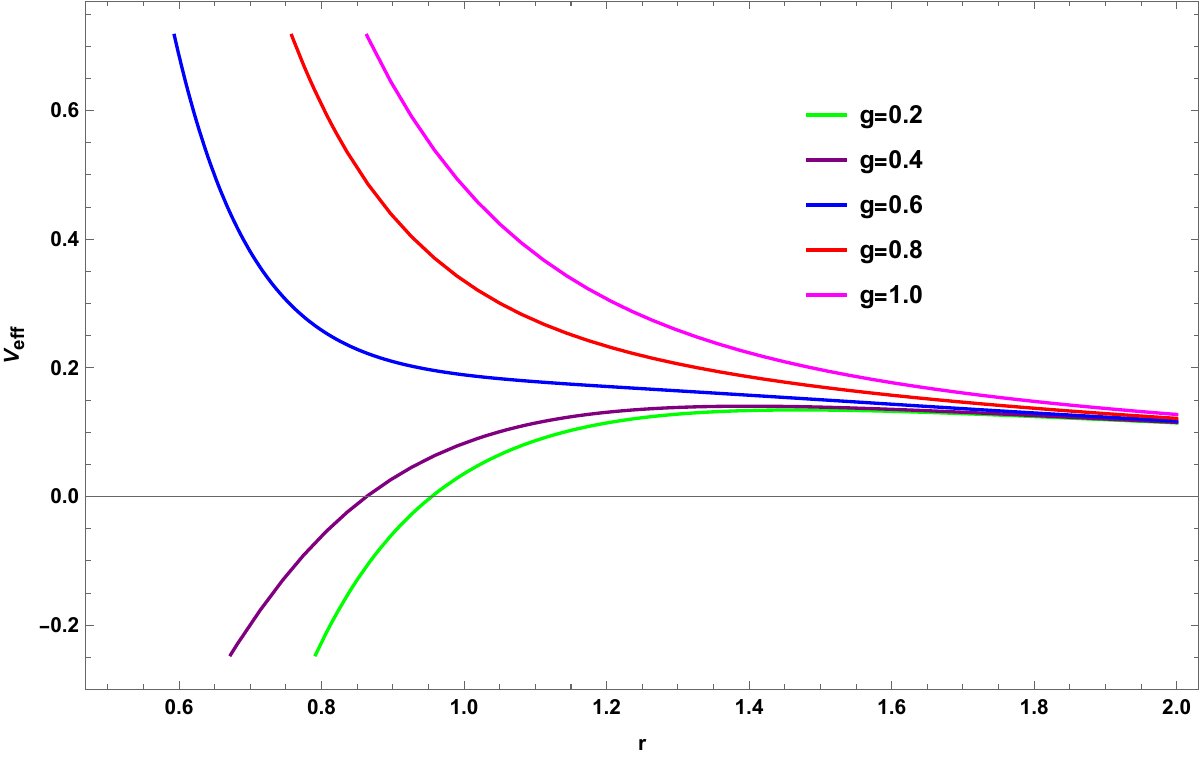}}\quad\quad
\subfloat[$g=0.1,L=1,\Lambda=-0.1,w=-2/3$]{\centering{}\includegraphics[scale=0.4]{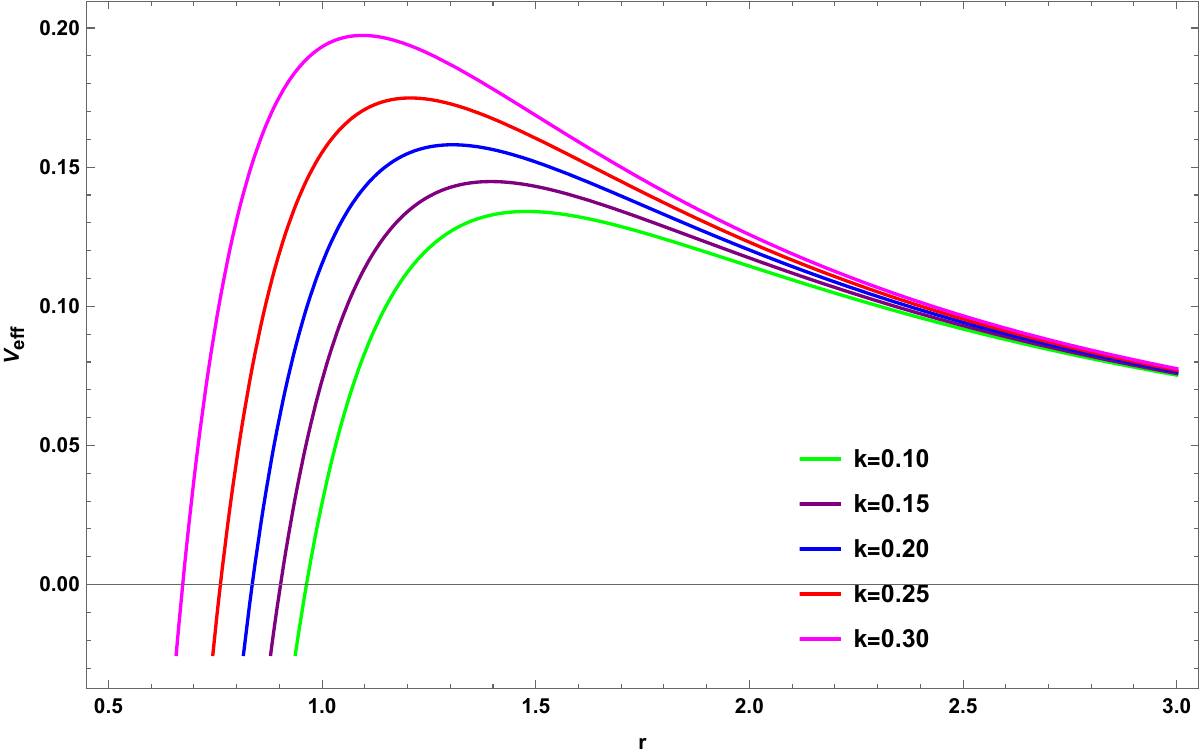}}
\hfill\\
\subfloat[$k=0.1=g,\Lambda=-0.3,w=-2/3$]{\centering{}\includegraphics[scale=0.4]{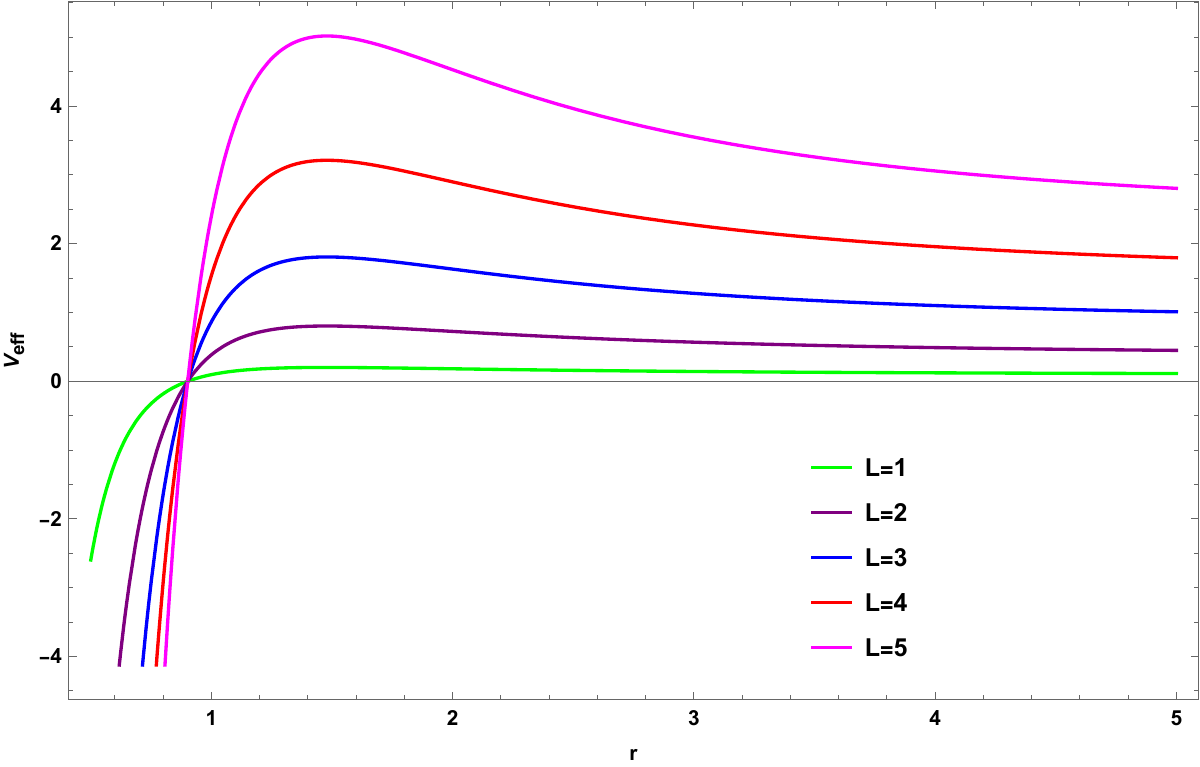}}\quad\quad
\subfloat[$g=0.1,k=0.001,L=1,w=-2/3$]{\centering{}\includegraphics[scale=0.4]{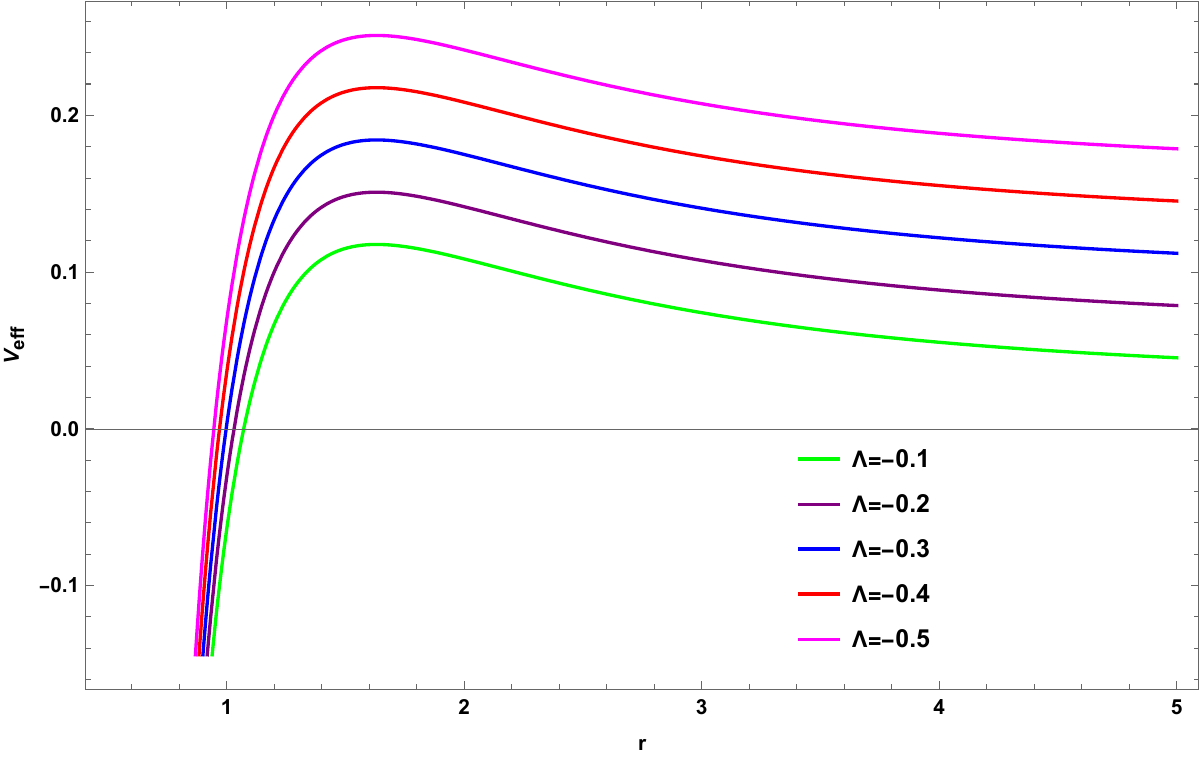}}
\hfill\\
\subfloat[$L=1,\Lambda=-0.1,w=-2/3$]{\centering{}\includegraphics[scale=0.4]{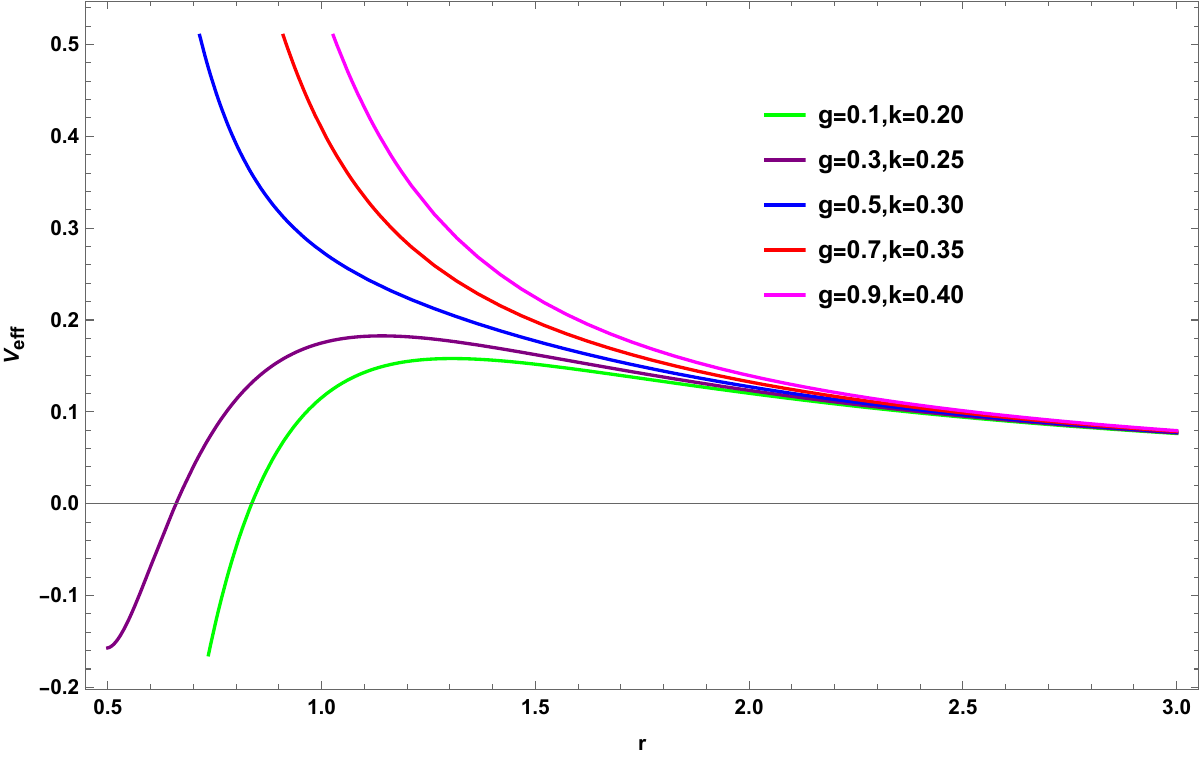}}\quad\quad
\subfloat[$g=0.1,k=0.001,L=1,\Lambda=-0.1$]{\centering{}\includegraphics[scale=0.4]{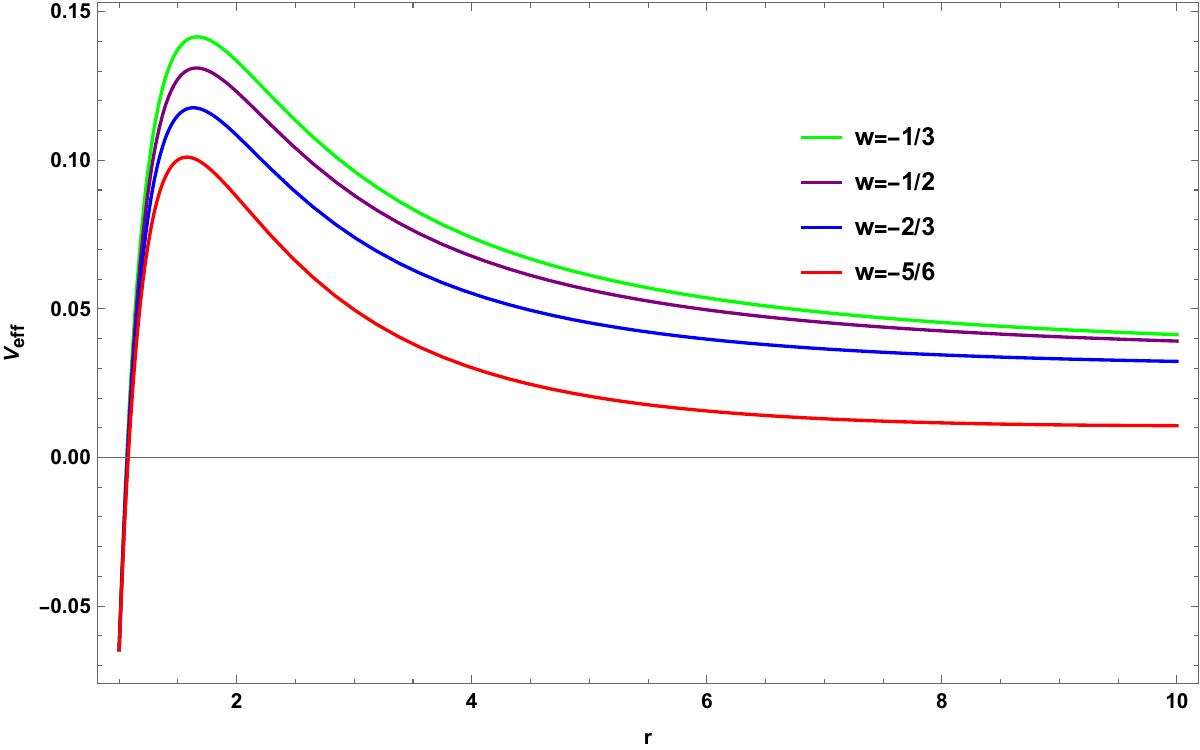}}
\centering{}\caption{The effective potential of null geodesics. Here we set $M=1/2, c=0.1$.}\label{fig:5}
\end{figure}
\par\end{center}

\begin{center}
\begin{figure}[ht!]
\subfloat[$k=0.1,L=1,\Lambda=-0.1,w=-2/3$]{\centering{}\includegraphics[scale=0.4]{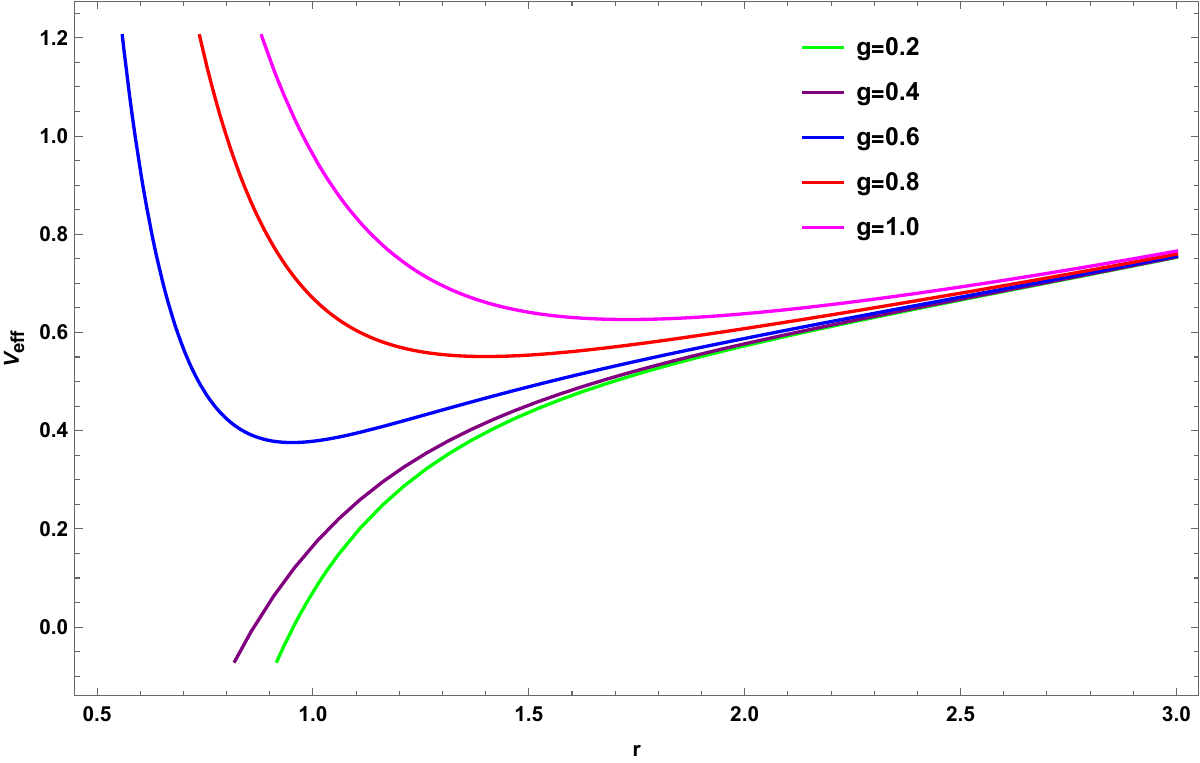}}\quad\quad
\subfloat[$g=0.1,L=1,\Lambda=-0.1,w=-2/3$]{\centering{}\includegraphics[scale=0.4]{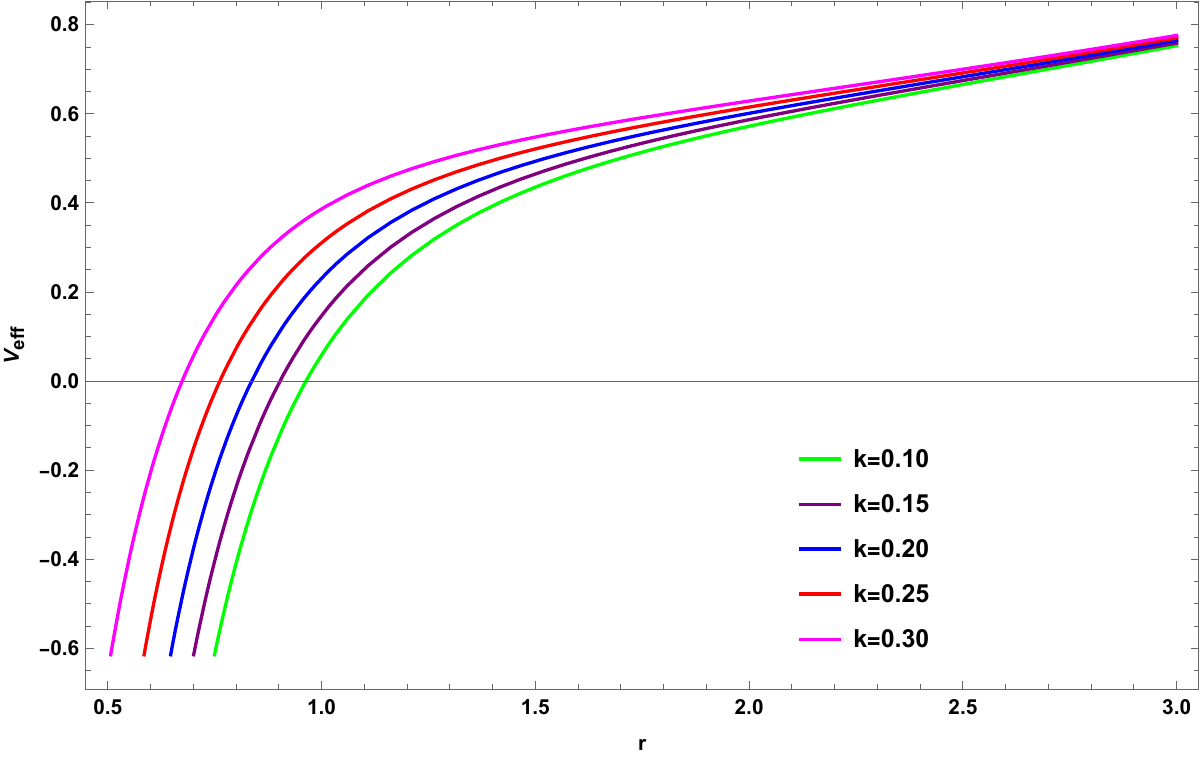}}
\hfill\\
\subfloat[$k=0.1=g,\Lambda=-0.3,w=-2/3$]{\centering{}\includegraphics[scale=0.4]{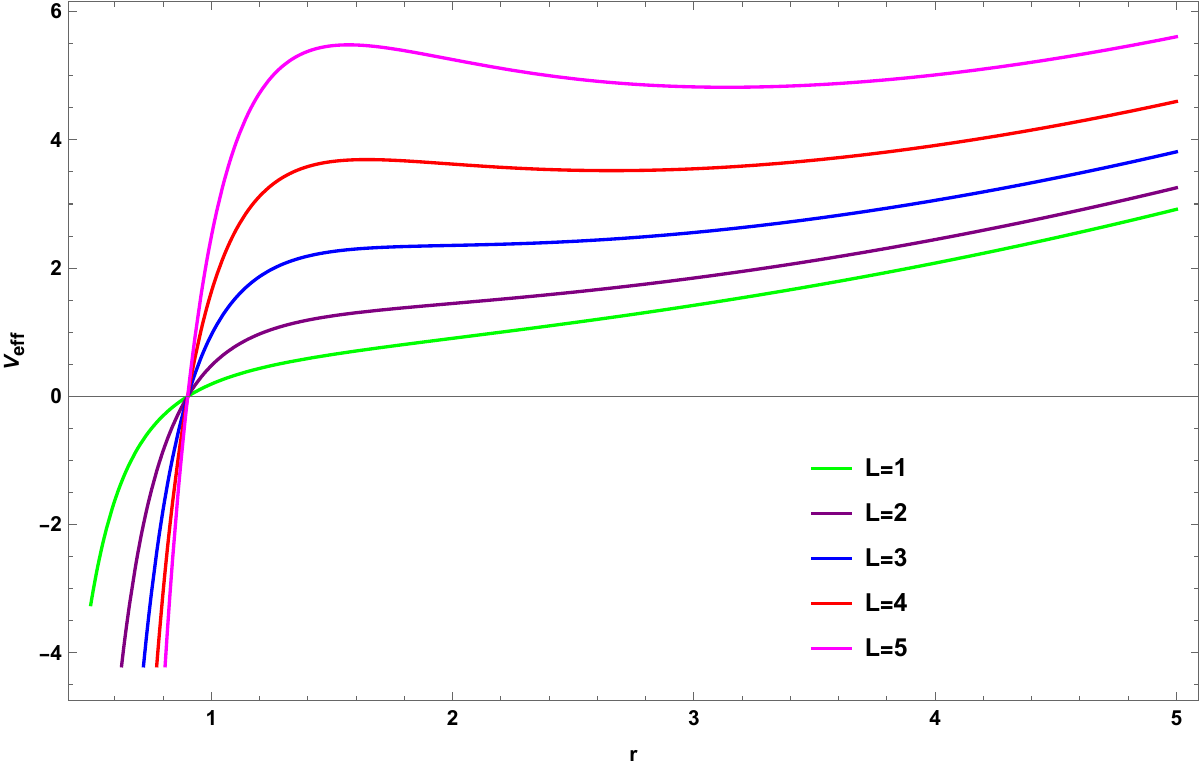}}\quad\quad
\subfloat[$g=0.1,k=0.001,L=1,w=-2/3$]{\centering{}\includegraphics[scale=0.4]{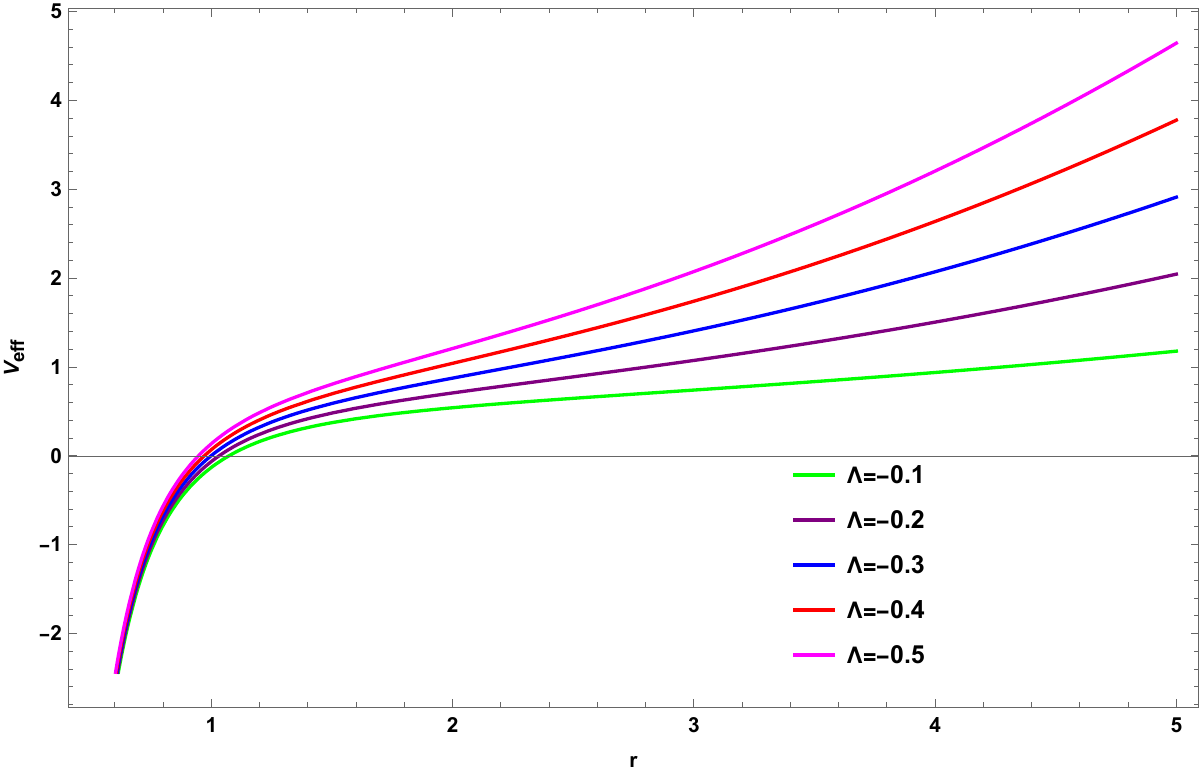}}
\hfill\\
\subfloat[$L=1,\Lambda=-0.1,w=-2/3$]{\centering{}\includegraphics[scale=0.4]{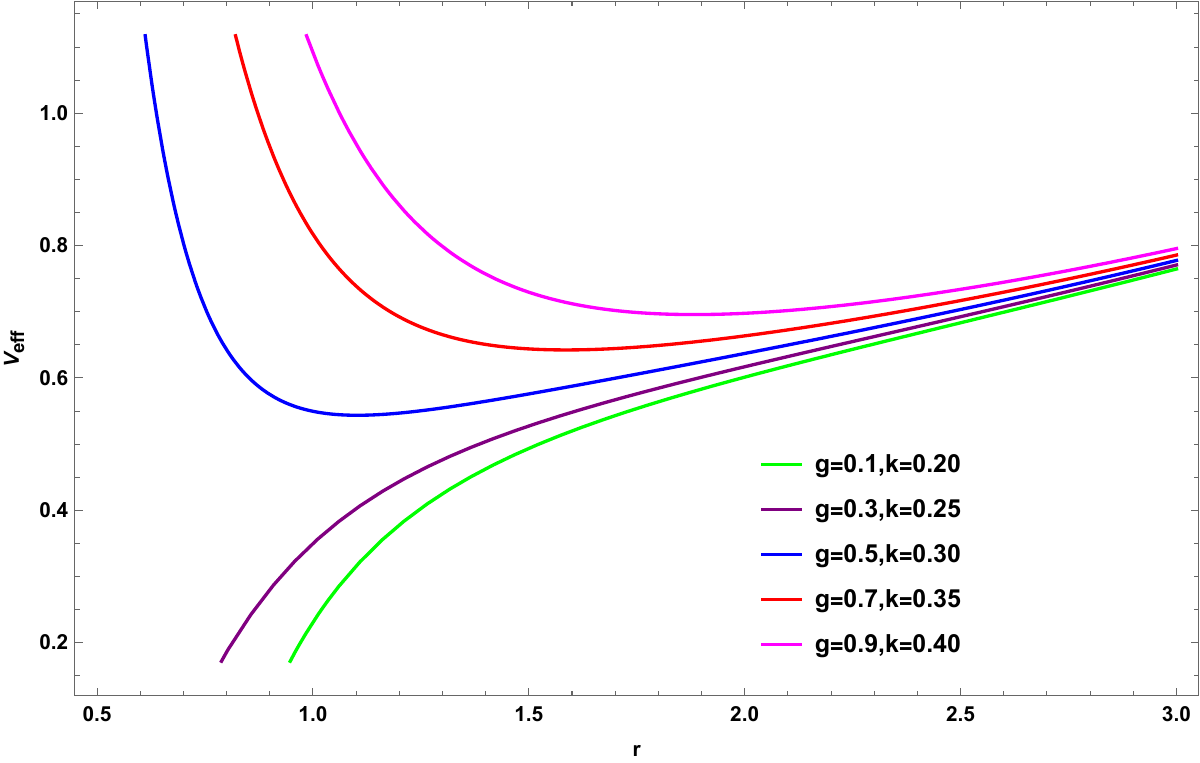}}\quad\quad
\subfloat[$g=0.5,k=0.2,L=1,\Lambda=-0.1$]{\centering{}\includegraphics[scale=0.4]{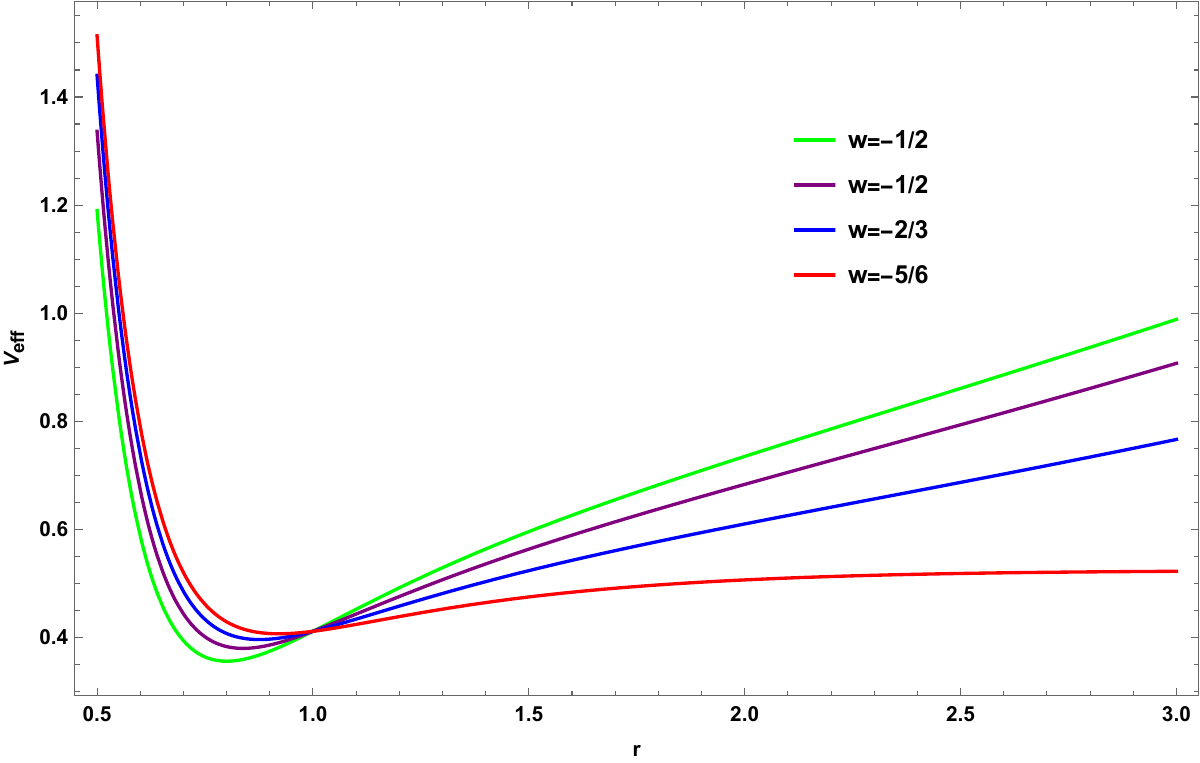}}
\centering{}\caption{The effective potential of time-like geodesics. Here we set $M=1/2, c=0.1$.}\label{fig:6}
\end{figure}
\par\end{center}

If we use Eqs. (\ref{cc2}) and (\ref{cc3}), we find the following one
\begin{equation}
    \left(\frac{dr}{d\phi}\right)^2=r^4\,\left[\frac{1}{\beta^2}-\frac{1}{\mathrm{L}^2}\,\left(-\epsilon+\frac{\mathrm{L}^2}{r^2}\right)\,\left(1-\frac{2\,M\,r^2}{r^3+g^3}\,e^{-k/r}-\frac{\Lambda\,r^2}{3}-\frac{c}{r^{3\,w+1}}\right)\right],\label{cc6}
\end{equation}
where $\beta=\mathrm{L}/\mathrm{E}$ is the impact parameter.

Now, we consider the motion of photon. For light-like geodesics, $\epsilon=0$, from Eq. (\ref{cc6}) 
\begin{equation}
    \left(\frac{dr}{d\phi}\right)^2=r^4\,\left[\frac{1}{\beta^2}-\frac{1}{r^2}\,\left(1-\frac{2\,M\,r^2}{r^3+g^3}\,e^{-k/r}-\frac{\Lambda\,r^2}{3}-\frac{c}{r^{3\,w+1}}\right)\right].\label{cc7}
\end{equation}

With the transformation $u=1/r$, Eq. (\ref{cc7}) becomes
\begin{equation}
    \left(\frac{du}{d\phi}\right)^2=\mathrm{G}(u,\beta,k,g,w,c),\label{cc10}
\end{equation}
where
\begin{equation}
    \mathrm{G}(u,\beta,k,g,w,c)=\frac{1}{\beta^2}-u^2\,\left(1-\frac{2\,M\,u}{1+u^3\,g^3}\,e^{-k\,u}-\frac{\Lambda}{3\,u^2}-c\,u^{3\,w+1}\right).\label{cc11}
\end{equation}
For photon, parameter $\beta$ is the distance from the black hole to the asymptotic line of photon orbit at infinity.

Defining a function as follow
\begin{equation}\label{cc12}
\Phi_\beta (r)=\int^{1/r}_{0}\,\frac{du}{\sqrt{\mathrm{G}(u,\beta,k,g,w,c)}}    
\end{equation}
into the equation (\ref{cc10}) and after simplification one can find the polar angle as,
\begin{eqnarray}\label{cc13}
\phi(\beta)=\left\{\,
    \begin{array}{cc}
        \Phi_{\beta}(r_h),\quad\quad \beta<\beta_c,\\
        2\,\Phi_{\beta} (r_m),\quad \beta>\beta_c.
    \end{array}
    \right.
\end{eqnarray}
In the limit $\beta \to \beta_c$, where $r=r_{ph}$ and $u \to u_{ph}=1/r_{ph}$, then the function $\mathrm{G}(u_{ph},\beta_c,k,g)=0$ which results that $\lim_{\beta \to \beta_c}\,\phi(\beta) \to +\infty$. Here $r_h$ is the radius of horizon, $r_{ph}$ is the radius of photon sphere and $r_m$ is the maximum root of the effective potential. 
\begin{center}
\begin{figure}[ht!]
\subfloat[$k=0.1,L=1,w=-2/3$]{\centering{}\includegraphics[scale=0.4]{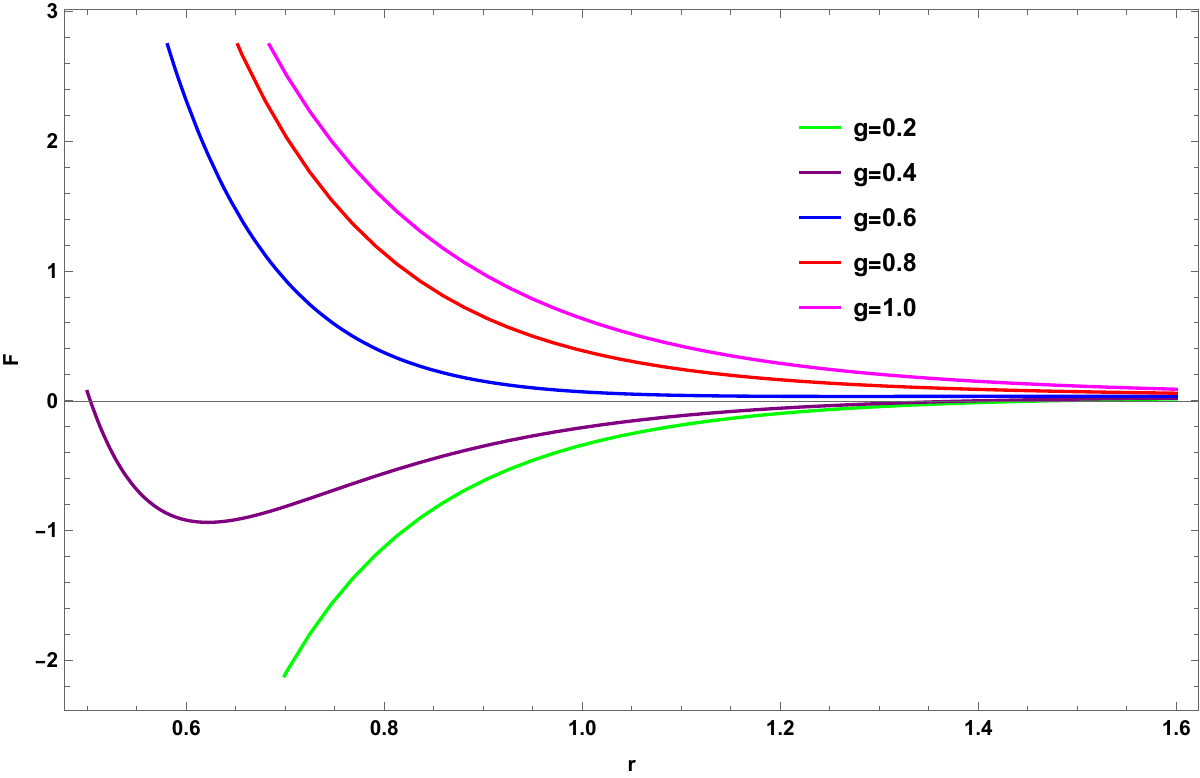}}\quad\quad
\subfloat[$g=0.1,L=1,w=-2/3$]{\centering{}\includegraphics[scale=0.4]{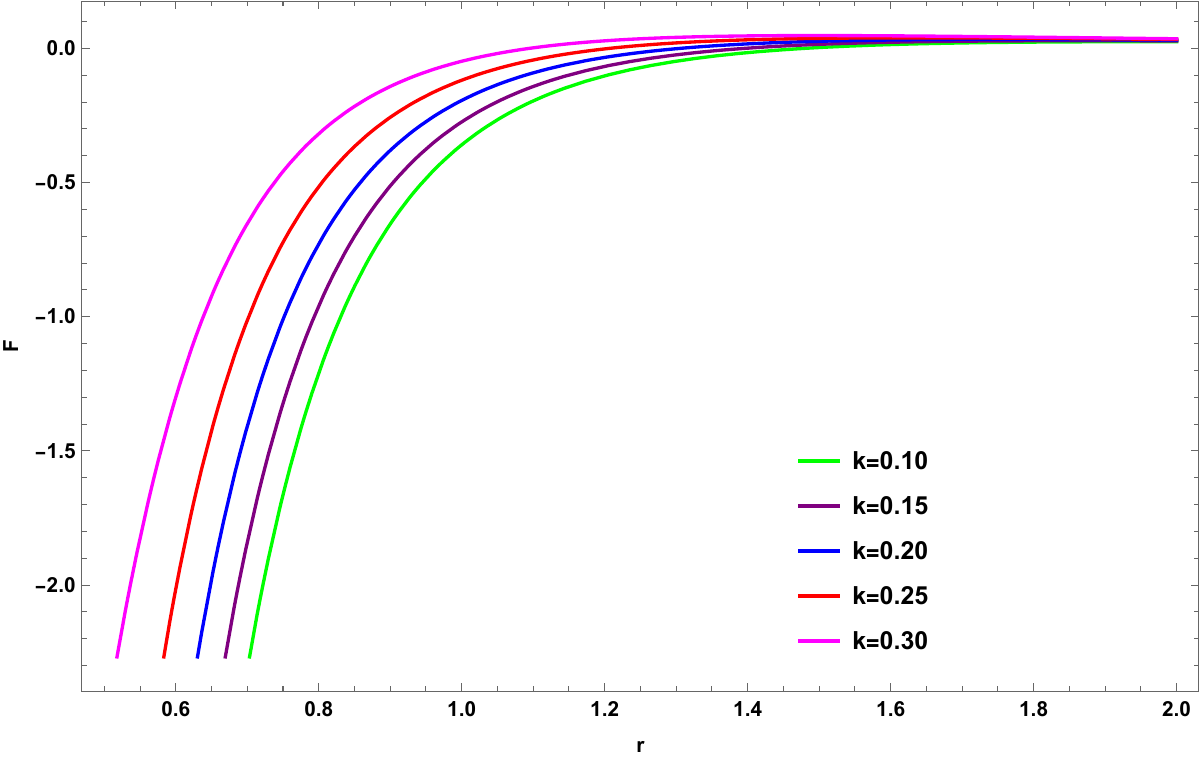}}
\hfill\\
\subfloat[$k=0.1=g,w=-2/3$]{\centering{}\includegraphics[scale=0.4]{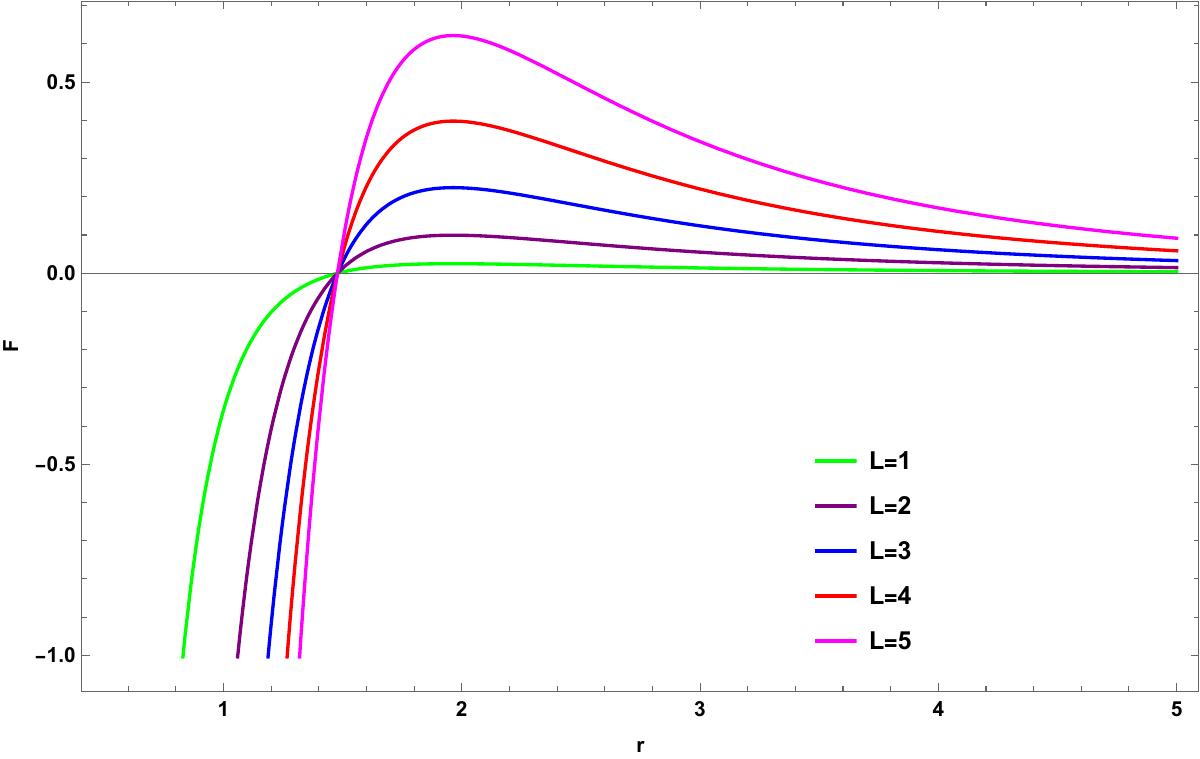}}\quad\quad
\subfloat[$L=1,w=-2/3$]{\centering{}\includegraphics[scale=0.4]{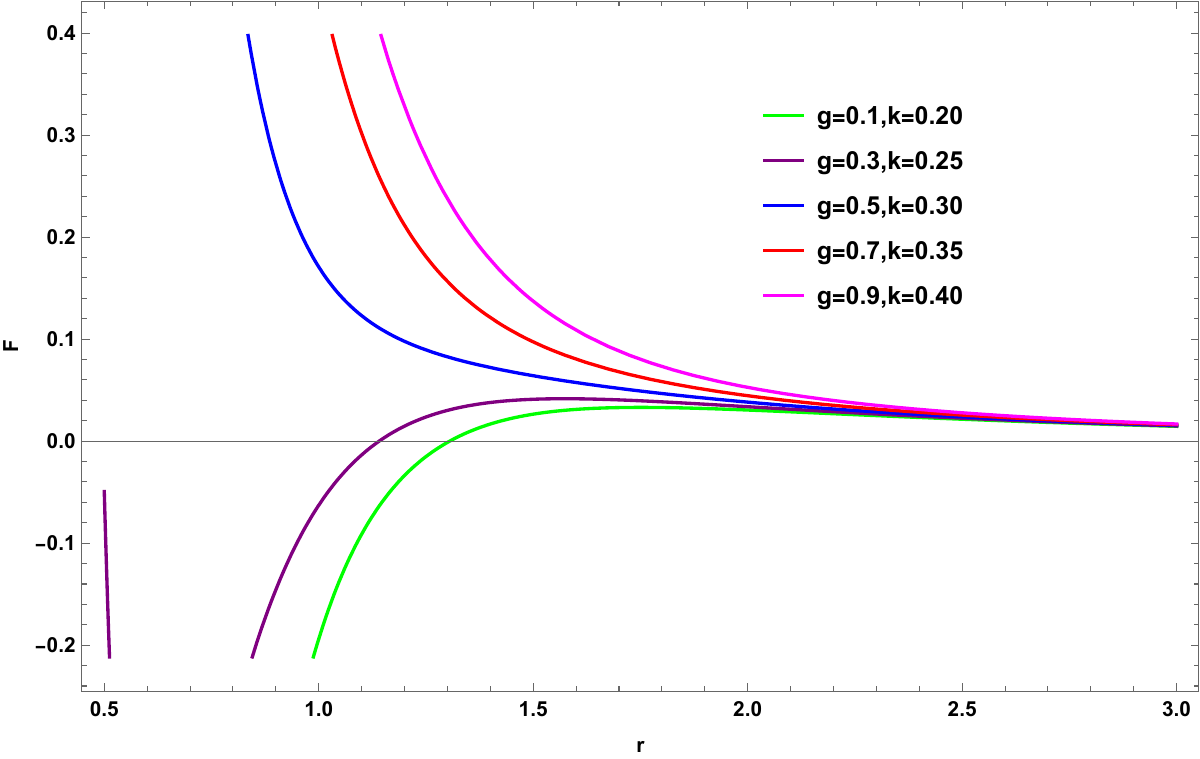}}
\hfill\\
\begin{centering}
\subfloat[$g=0.5,k=0.1,L=1$]{\centering{}\includegraphics[scale=0.4]{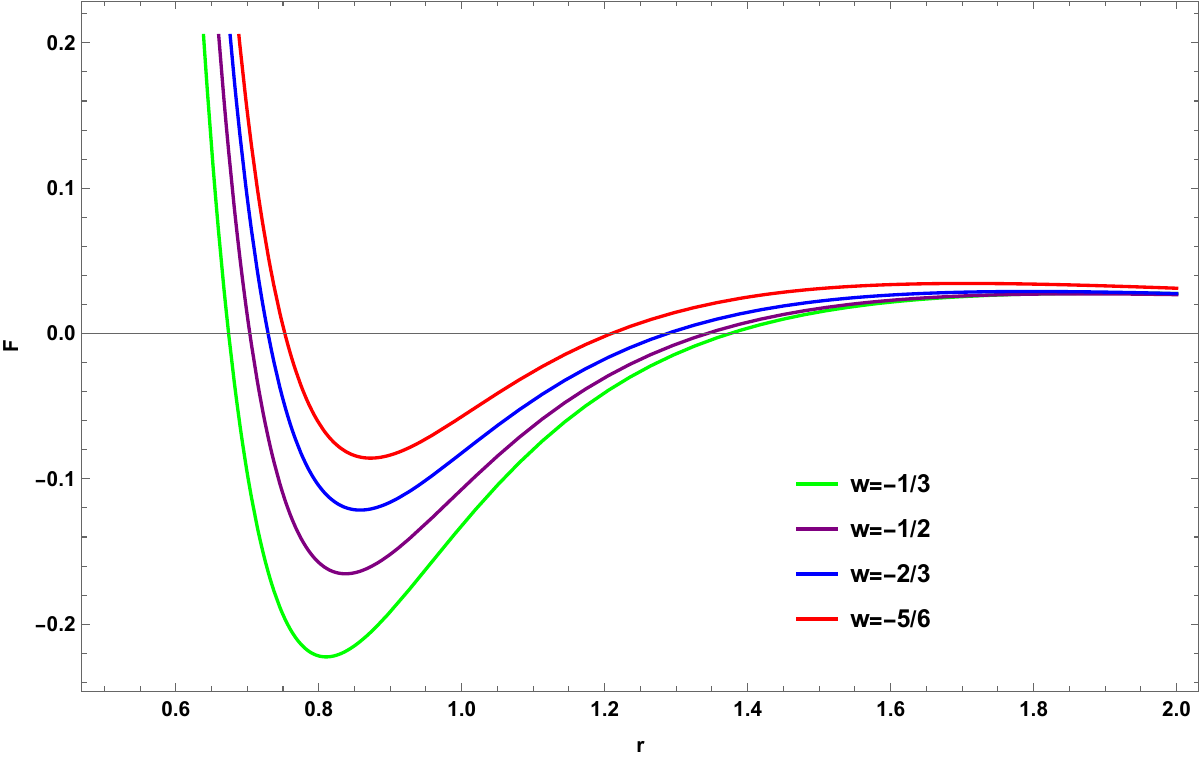}}
\par\end{centering}
\centering{}\caption{The force on photon particle. Here we set $M=1/2, c=0.1$.}\label{fig:8}
\end{figure}
\par\end{center}

\subsection{Force on the massless particle}

In this part, we try to calculate the force acting on the massless photon particles and analyze the result how NLED and QF affect this force for a particular state parameter.

The force on the massless photon particle can be obtained by using the effective potential $V_{eff}$ as,
\begin{equation*}
\mathrm{F}(r)=-\frac{1}{2}\frac{dV_\text{eff}\left( r\right) }{dr}
\end{equation*} 

Using the effective potential given in Eq. (\ref{cc4}) for the photon particle, we find the following expression of this force given by
\begin{eqnarray}
    \mathrm{F}(r)&=&\frac{\mathrm{L}^2}{2\,r^2}\,\Bigg[\frac{2}{r}-\frac{6\,M\,r^4\,e^{-k/r}}{(g^3 + r^3)^2}-\frac{2\,k\,M\,e^{-k/r}}{
       g^3 + r^3}-c\,(1-3\,w)\,r^{-2-3\,w}\Bigg].\label{force}
\end{eqnarray}
The above force clearly shows that both parameters $(k,g)$, along with the conserved angular momentum $\mathrm{L}$, influence the force. As a result, this modifies the outcome compared to the known expression for the AdS-Schwarzschild-Kiselev black hole with quintessence.

In Figure \ref{fig:8}, the force given by equation (\ref{force}) is plotted as a function of $r$ for various values of the parameters $k, g, \mathrm{L}$ and $w$. As can be observed from the expression for the force (\ref{force}), the first term represents a repulsive force, while the second to fourth terms contribute to the attractive nature of the overall force. This attractive behavior is further influenced by the state parameter $w$, which lies within the range $-1 < w < -\frac{1}{3}$. The value of $w$ governs the strength and nature of the attractive force, with its influence becoming more pronounced as $w$ moves further into this range. In this figure, we observe that depending on the values of $k$, $g$ and $\mathrm{L}$, the force can exhibit both attractive and repulsive characteristics. The nature of the force changes as these parameters vary, illustrating the complex interplay between them.

For $k=0$, we find form Eq. (\ref{force}) the force
\begin{eqnarray}
    \mathrm{F}(r)=\frac{\mathrm{L}^2}{2\,r^2}\,\Bigg[\frac{2}{r}-
    -\frac{6\,M\,r^4}{(g^3 + r^3)^2}-3\,c\,(1+w)\,r^{-2-3\,w}\Bigg].\label{force2}
\end{eqnarray}
which is the similar to the result for Bardeen-Kiselev black hole having state parameter $w$.

For $k=0=g$, we find this force
\begin{eqnarray}
    \mathrm{F}(r)=\frac{\mathrm{L}^2}{2\,r^2}\,\left[\frac{2}{r}-\frac{6\,M}{r^2}-3\,c\,(1+w)\,r^{-2-3\,w}\right].\label{force3}
\end{eqnarray}
which is the exact expression of force on the photon particle for Schwarzschild black hole with quintessence matter having state parameter $w$.

\section{The Regge-Wheeler Potential}\label{sec:5}

In this part, we investigate the spin-dependent Regge-Wheeler potentials for the regular black hole constructed in this paper. The Regge-Wheeler-Zerilli equations describe the gravitational perturbations of a Schwarzschild-like black hole in general relativity. These perturbations are categorized into two types: axial and polar perturbations. The equation governing axial perturbations is known as the Regge-Wheeler equation \cite{TR}, while the equation for polar perturbations is referred to as the Zerilli equation \cite{FJZ}. Axial perturbations are typically more complex and do not lend themselves easily to the WKB approximation, making the computation of quasi-normal modes (QNMs) challenging without the use of numerical methods. Given this difficulty, we only examine the relevant Regge-Wheeler potential for fields, such as zero-spin, spin-one and spin-two and analyze the result. 

To proceed for the RW-potential, we perform the following coordinate change (called tortoise coordinate) 
\begin{equation}
    dr_*=\frac{dr}{f(r)},\quad\quad \partial_{r_*}=f(r)\,\partial_r.\label{q2}
\end{equation}
into the line-element Eq. (\ref{qq1}) results
\begin{equation}
    ds^2=\mathcal{F}(r_*)\,\{-dt^2+dr^2_{*}\}+\mathcal{H}(r_*)\,(d\theta^2+\sin^2 \theta\,d\phi^2),\label{qq3}
\end{equation}

In Regge and Wheeler’s original work \cite{TR}, they show that for perturbations in a black hole space-time, assuming a separable wave form of the type
\begin{equation}
    \Phi(t, r_{*},\theta, \phi)=\exp(i\,\omega\,t)\,\psi(r_*)\,Y^{\ell}_{m} (\theta,\phi),\label{qq4}
\end{equation}
where $Y^{\ell}_{m} (\theta,\phi)$ are the spherical harmonics, $\omega$ is (possibly complex) temporal frequency in the Fourier domain \cite{TR}, and $\psi (r)$ is a propagating scalar, vector, or spin two axial bi-vector field in the candidate space-time. The Regge-Wheeler equation is given by
\begin{equation}
    \frac{\partial^2 \psi(r_*)}{\partial r^2_{*}}+\left\{\omega^2-\mathcal{V}_s\right\}\,\psi(r_*)=0.\label{qq5}
\end{equation}
The method for solving Equation (\ref{qq5}) is dependent on the spin of the perturbations and on the background space-time.

The spin-dependent Regge-Wheeler potential is given by the following expression
\begin{equation}
    \mathcal{V}_S=\frac{\mathcal{F}}{\mathcal{H}^2}\,\Big\{\ell\,(\ell+1)+S\,(S-1)\,(g^{rr}-1)\Big\}+(1-S)\,\frac{\partial^2_{r_{*}}\,\mathcal{H}}{\mathcal{H}},\label{qq6}
\end{equation}
where $\mathcal{F}$ and $\mathcal{H}$ are the relevant functions as specified by Equation (\ref{qq1}), $\ell$ is the multipole number $\ell\geq s$, and $g^{rr}$ is the relevant contrametric component with respect to standard curvature coordinates (for which the covariant components are presented in Equation (\ref{qq1})).

For the metric (\ref{qq1}) under consideration, we have
\begin{eqnarray}
    &&\mathcal{F}(r)=1-\frac{2\,M\,r^2}{r^3+g^3}\,e^{-k/r}-\frac{\Lambda}{3}\,r^2-\frac{c}{r^{3\,w+1}},\quad\quad \mathcal{H}(r)=r,\nonumber\\
    &&g^{rr}=1-\frac{2\,M\,r^2}{r^3+g^3}\,e^{-k/r}-\frac{\Lambda}{3}\,r^2-\frac{c}{r^{3\,w+1}}.\label{qq7}
\end{eqnarray}
Hence, we find
\begin{equation}
    \frac{\partial^2_{r_{*}}\,\mathcal{H}}{\mathcal{H}}=\frac{f(r)\,f'(r)}{r}.\label{qq8}
\end{equation}

So the spin-dependent RW potential is given by

\begin{eqnarray}
    \mathcal{V}_S&=&\left(1-\frac{2Mr^2}{r^3+g^3}e^{-k/r}-\frac{\Lambda}{3}r^2-\frac{c}{r^{3\,w+1}}\right)\Bigg[\frac{\ell(\ell+1)}{r^2}+\frac{S(S-1)}{r^2}\Bigg\{-\frac{2Mr^2}{r^3+g^3}e^{-k/r}-\frac{\Lambda}{3}r^2-\frac{c}{r^{3\,w+1}}\Bigg\}\Bigg]\nonumber\\
    &+&\frac{(1-S)}{r}\,\left(1-\frac{2\,M\,r^2}{r^3+g^3}\,e^{-k/r}-\frac{\Lambda}{3}\,r^2-\frac{c}{r^{3\,w+1}}\right)\times\nonumber\\
    &&\Bigg(\frac{6\,e^{-k/r}\,M\,r^4}{(g^3 + r^3)^2}-\frac{2\,e^{-k/r}\,k\,M}{
       g^3 + r^3}-\frac{4\,e^{-k/r}\,M\,r}{g^3+r^3}- 
       c\,(-1-3\,w)\,r^{-2-3\,w}\Bigg).\label{qq9}
\end{eqnarray}

 From the spin-dependent RW potential expression (\ref{qq9}), it is evident that several parameters influence the potential. These include the deviation parameter $k$, the Hayward-like parameter $g$, the cosmological constant $\Lambda$, the state parameter $w$, and the multipole number $\ell$. Below, we discuss some special cases of the above result. 

In the limit where $k=0$, this RW-potential (\ref{qq9}) reduces to as follows:
\begin{eqnarray}
    \mathcal{V}_S&=&\left(1-\frac{2\,M\,r^2}{r^3+g^3}-\frac{\Lambda}{3}\,r^2-\frac{c}{r^{3\,w+1}}\right)\Bigg[\frac{\ell\,(\ell+1)}{r^2}+\frac{S\,(S-1)}{r^2}\Bigg\{-\frac{2\,M\,r^2}{r^3+g^3}-\frac{\Lambda}{3}\,r^2-\frac{c}{r^{3\,w+1}}\Bigg\}\Bigg]\nonumber\\
    &+&\frac{(1-S)}{r}\,\left(1-\frac{2\,M\,r^2}{r^3+g^3}-\frac{\Lambda}{3}\,r^2-\frac{c}{r^{3\,w+1}}\right)
    \Bigg[\frac{6\,M\,r^4}{(g^3 + r^3)^2}-\frac{4\,M\,r}{g^3+r^3}- 
       c\,(-1-3\,w)\,r^{-2-3\,w}\Bigg].\label{qq9a}
\end{eqnarray}
Equation (\ref{qq9a}) is the spin-dependent RW-potential for Hayward black hole metric with quintessence field featuring a negative cosmological constant.

In the limit where $k=0=g$, we find from Eq. (\ref{qq9}) 
\begin{eqnarray}
    \mathcal{V}_S&=&\left(1-\frac{2\,M}{r}-\frac{\Lambda}{3}r^2-\frac{c}{r^{3\,w+1}}\right)\,\Bigg[\frac{\ell(\ell+1)}{r^2}+\frac{S(S-1)}{r^2}\Bigg\{-\frac{2\,M}{r}-\frac{\Lambda}{3}r^2-\frac{c}{r^{3\,w+1}}\Bigg\}\Bigg]\nonumber\\
    &+&\frac{(1-S)}{r}\,\left(1-\frac{2\,M}{r}-\frac{\Lambda}{3}\,r^2-\frac{c}{r^{3\,w+1}}\right)
    \left(\frac{2\,M}{r^2}- 
       c\,(-1-3\,w)\,r^{-2-3\,w}\right).\label{qq9b}
\end{eqnarray}
Equation (\ref{qq9b}) is the spin-dependent RW-potential for Kiselev black hole metric featuring a negative cosmological constant.

Additionally, in the limit where $k=0$ and $c=0$, we find from Eq. (\ref{qq9}) 
\begin{eqnarray}
    \mathcal{V}_S&=&\left(1-\frac{2\,M\,r^2}{r^3+g^3}-\frac{\Lambda}{3}r^2\right)\left[\frac{\ell(\ell+1)}{r^2}+\frac{S(S-1)}{r^2}\left\{-\frac{2\,M\,r^2}{r^3+g^3}-\frac{\Lambda}{3}r^2\right\}\right]\nonumber\\
    &+&\frac{(1-S)}{r}\,\left(1-\frac{2\,M\,r^2}{r^3+g^3}-\frac{\Lambda}{3}\,r^2\right)\left(\frac{6\,M\,r^4}{(g^3 + r^3)^2}-\frac{4\,M\,r}{g^3+r^3}\right).\label{qq9c}
\end{eqnarray}
Equation (\ref{qq9c}) is the spin-dependent RW-potential for Hayward-like black hole metric featuring a negative cosmological constant.

Moreover, in the limit where $k=0=g$ and $c=0$, we find from Eq. (\ref{qq9}) 
\begin{eqnarray}
    \mathcal{V}_S&=&\left(1-\frac{2\,M}{r}-\frac{\Lambda}{3}\,r^2\right)\left[\frac{\ell(\ell+1)}{r^2}+\frac{S(S-1)}{r^2}\left\{-\frac{2\,M}{r}-\frac{\Lambda}{3}\,r^2\right\}
    +\frac{2\,M\,(1-S)}{r^3}\right].\label{qq9d}
\end{eqnarray}
Equation (\ref{qq9d}) is the spin-dependent RW-potential for Schwarzschild black hole metric featuring a negative cosmological constant.

\begin{center}
\begin{figure}[ht!]
\subfloat[$k=0.1,\ell=1,\Lambda=-0.1,w=-2/3$]{\centering{}\includegraphics[scale=0.4]{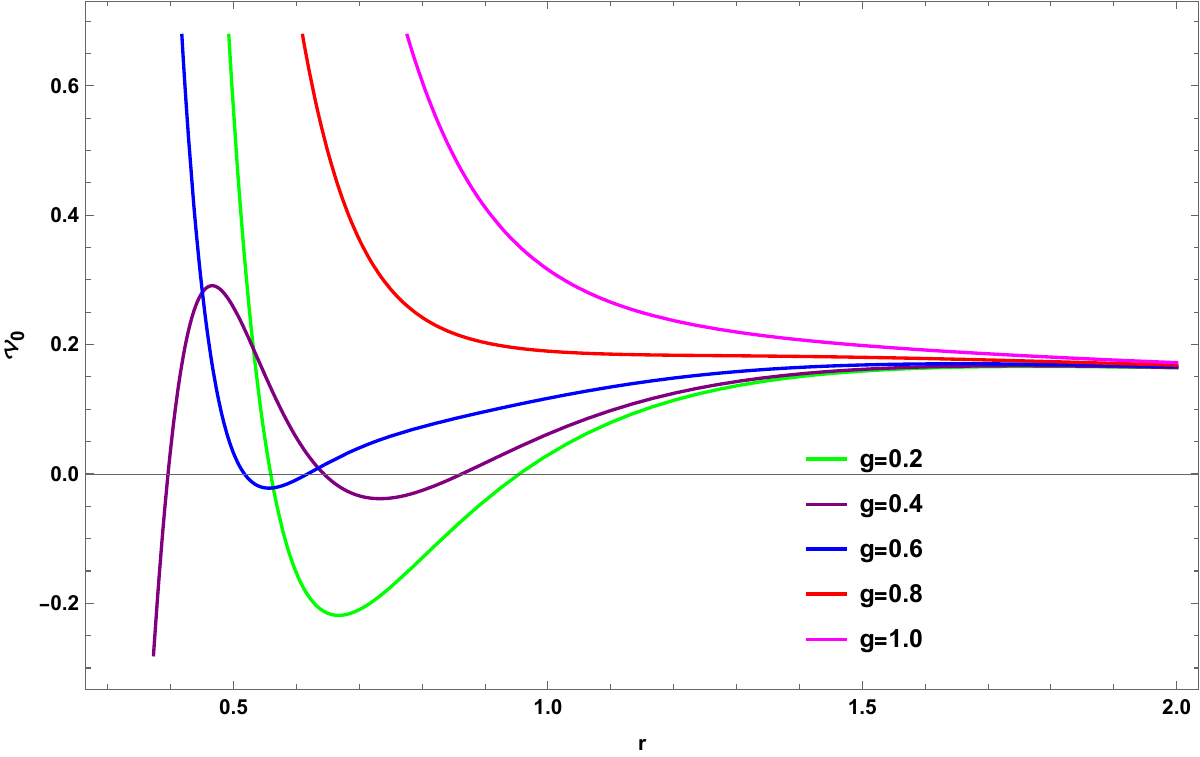}}\quad\quad
\subfloat[$g=0.1,\ell=1,\Lambda=-0.1,w=-2/3$]{\centering{}\includegraphics[scale=0.4]{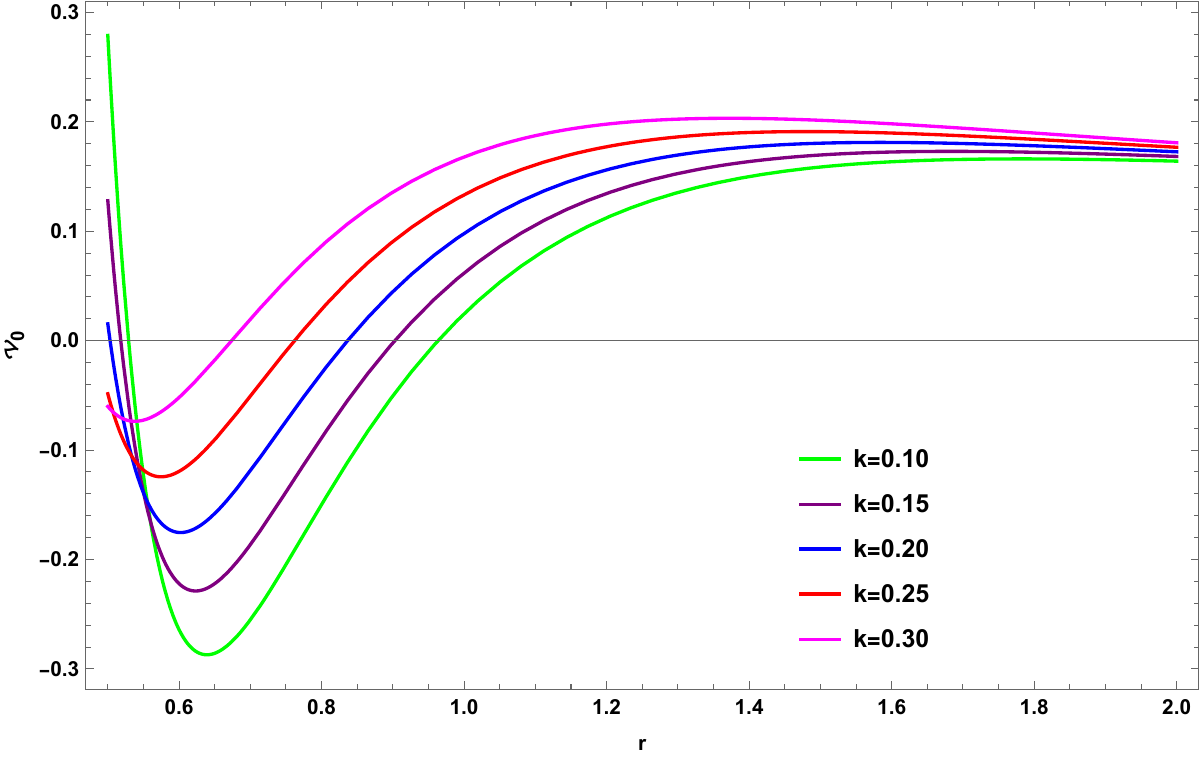}}
\hfill\\
\subfloat[$k=0.1=g,\Lambda=-0.3,w=-2/3$]{\centering{}\includegraphics[scale=0.4]{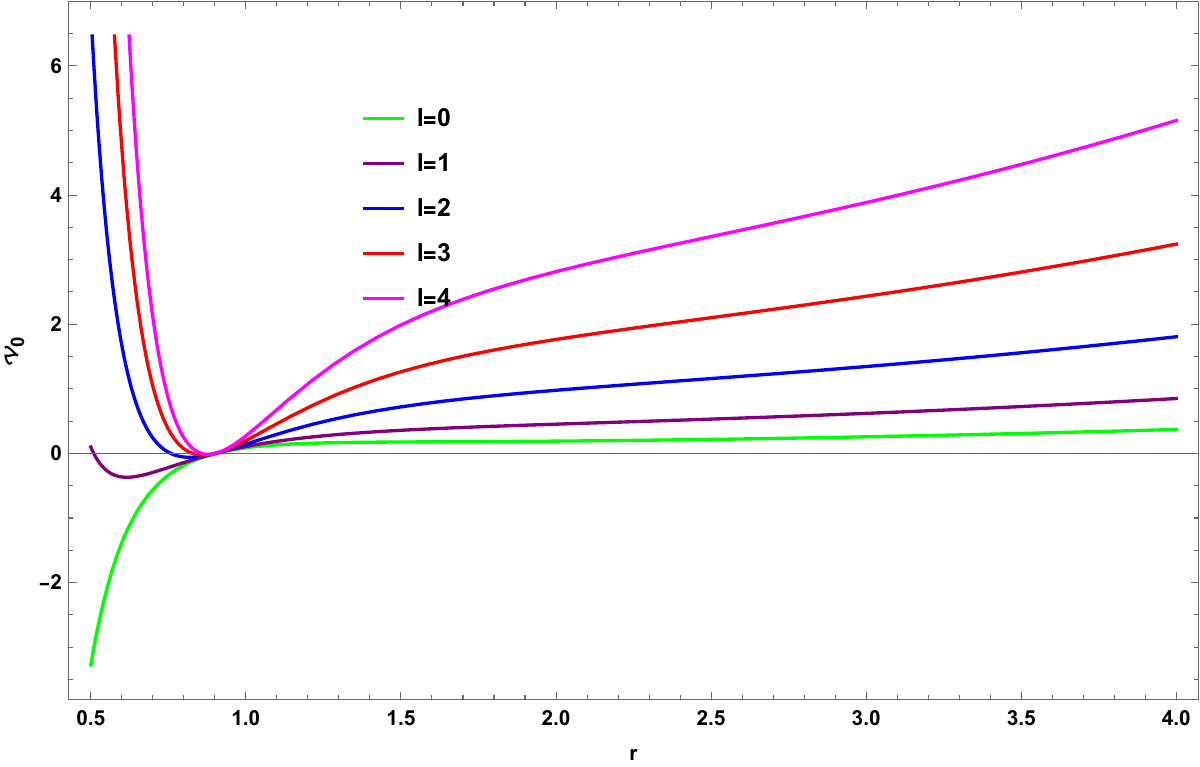}}\quad\quad
\subfloat[$g=0.1=k,\ell=1,w=-2/3$]{\centering{}\includegraphics[scale=0.4]{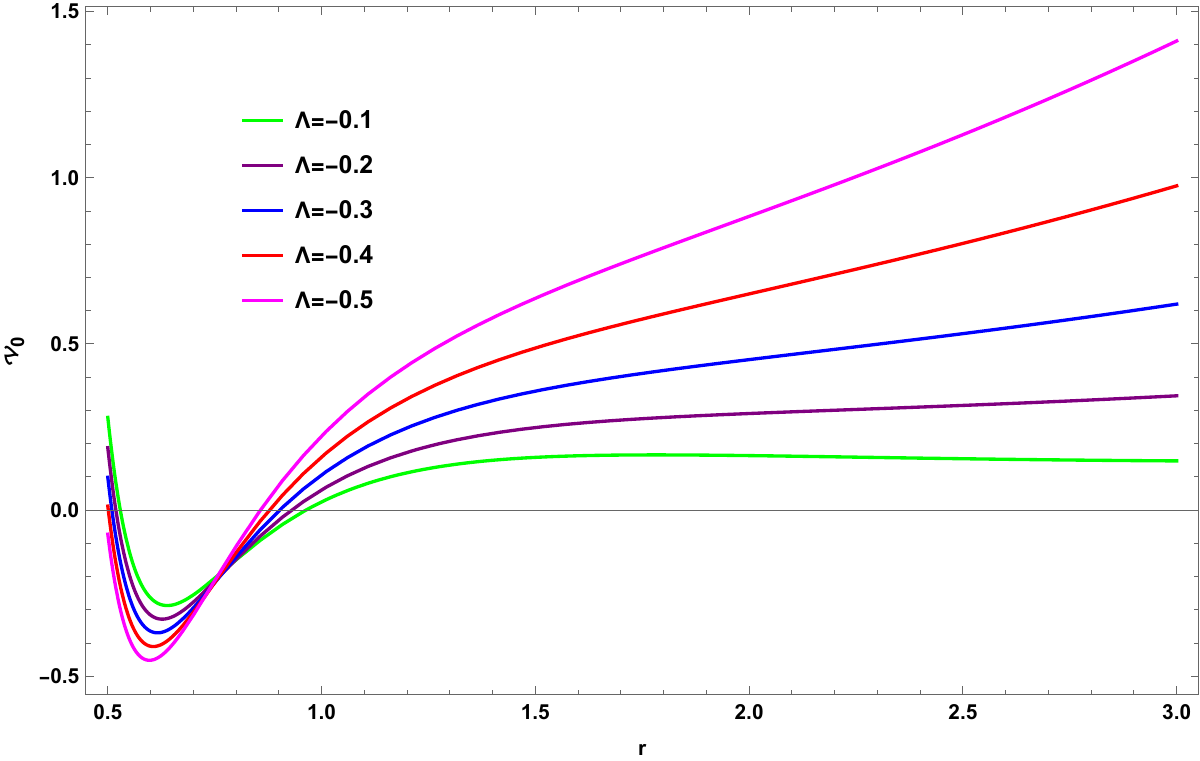}}
\hfill\\
\subfloat[$\ell=1,\Lambda=-0.1,w=-2/3$]{\centering{}\includegraphics[scale=0.4]{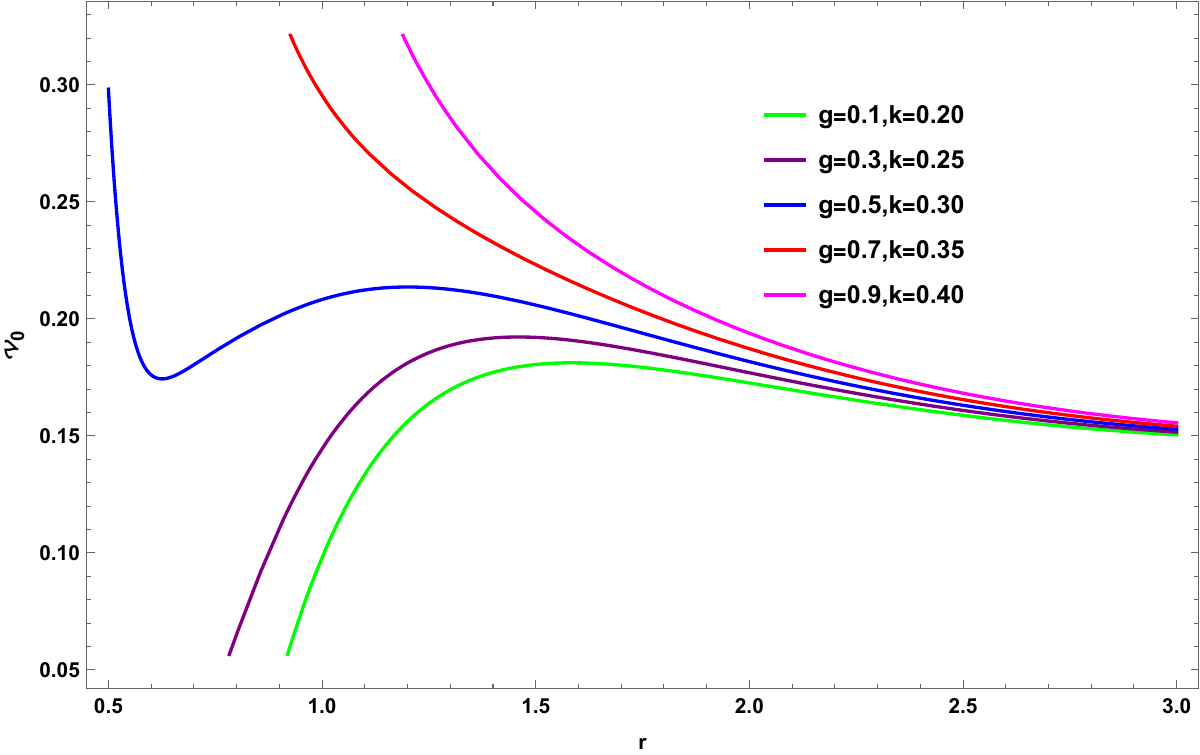}}\quad\quad
\subfloat[$g=0.5,k=0.1,\ell=1,\Lambda=-0.1$]{\centering{}\includegraphics[scale=0.4]{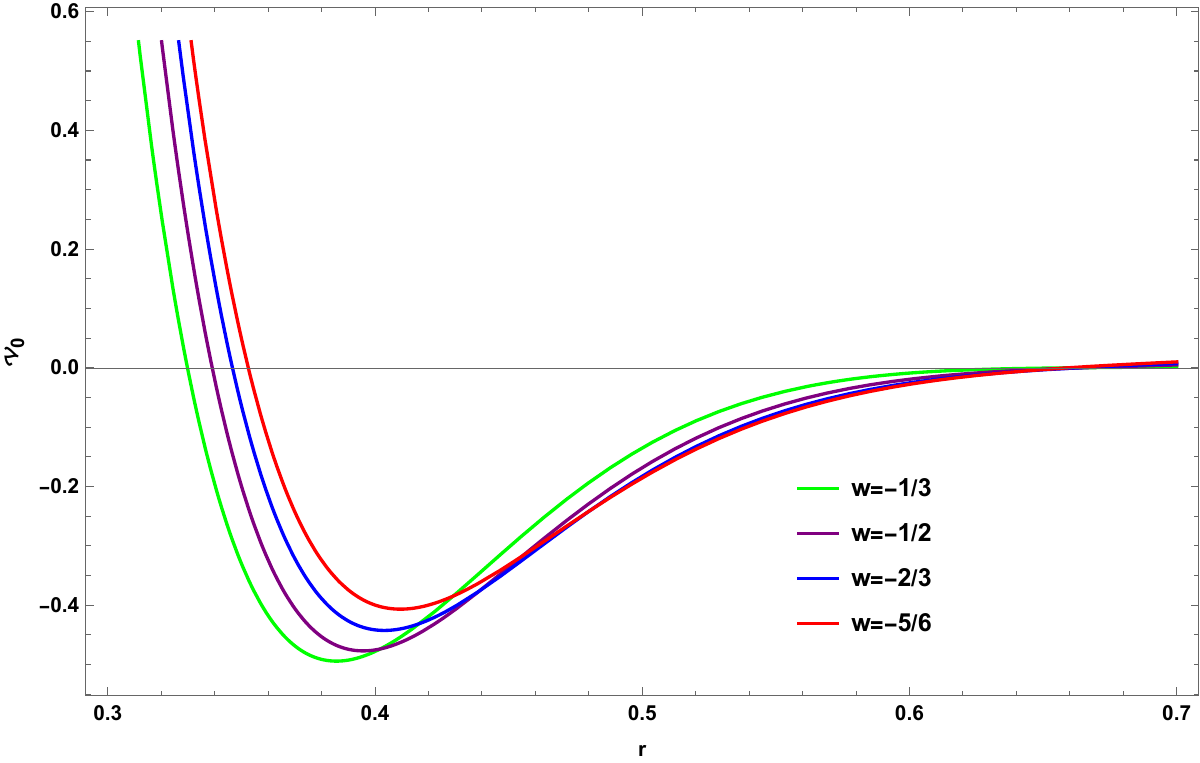}}
\centering{}\caption{The RW-potential of Eq. (\ref{qq10}) for zero spin scalar field with multipole number $\ell=1$. Here we set $M=1/2, c=0.1$.}\label{fig:9}
\end{figure}
\par\end{center}

Now, we discuss this RW potential (\ref{qq9}) for various spin fields, such as spin-0 scalar, spin-1 vector, and spin-2 tensor field and analyze the outcomes.

\begin{center}
    {\bf Case A: Zero spin field, $S=0$: Scalar Perturbations}
\end{center}

For zero spin, $S=0$ which corresponds to scalar perturbations, the RW-potential (\ref{qq9}) becomes
\begin{eqnarray}
    \mathcal{V}_0&=&\left(1-\frac{2\,M\,r^2}{r^3+g^3}\,e^{-k/r}-\frac{\Lambda}{3}\,r^2-\frac{c}{r^{3\,w+1}}\right)\,\frac{\ell\,(\ell+1)}{r^2}
    +\frac{1}{r}\,\left(1-\frac{2\,M\,r^2}{r^3+g^3}\,e^{-k/r}-\frac{\Lambda}{3}\,r^2-\frac{c}{r^{3\,w+1}}\right)\times\nonumber\\
    &&\Bigg(\frac{6\,e^{-k/r}\,M\,r^4}{(g^3 + r^3)^2}-\frac{2\,e^{-k/r}\,k\,M}{
       g^3 + r^3}-\frac{4\,e^{-k/r}\,M\,r}{g^3+r^3}- 
       c\,(-1-3\,w)\,r^{-2-3\,w}\Bigg).\label{qq10}
\end{eqnarray}
which reduced to  the scalar potential for the Hayward-Kiselev-AdS black hole metric provided $k=0$; Kiselev-AdS black hole metric provided $k=0=g$; Hayward-AdS black hole metric provided $k=0$ and $c=0$; and finally Schwarzschild-AdS black hole metric provided $k=0=g$ and $c=0$. 

We have generated a few graphs in Figure (\ref{fig:9}) of this RW-potential (\ref{qq10}) as a function of $r$ for different values other parameters and shown their behavior.

\begin{center}
\begin{figure}[ht!]
\subfloat[$k=0.2,w=-2/3$]{\centering{}\includegraphics[scale=0.4]{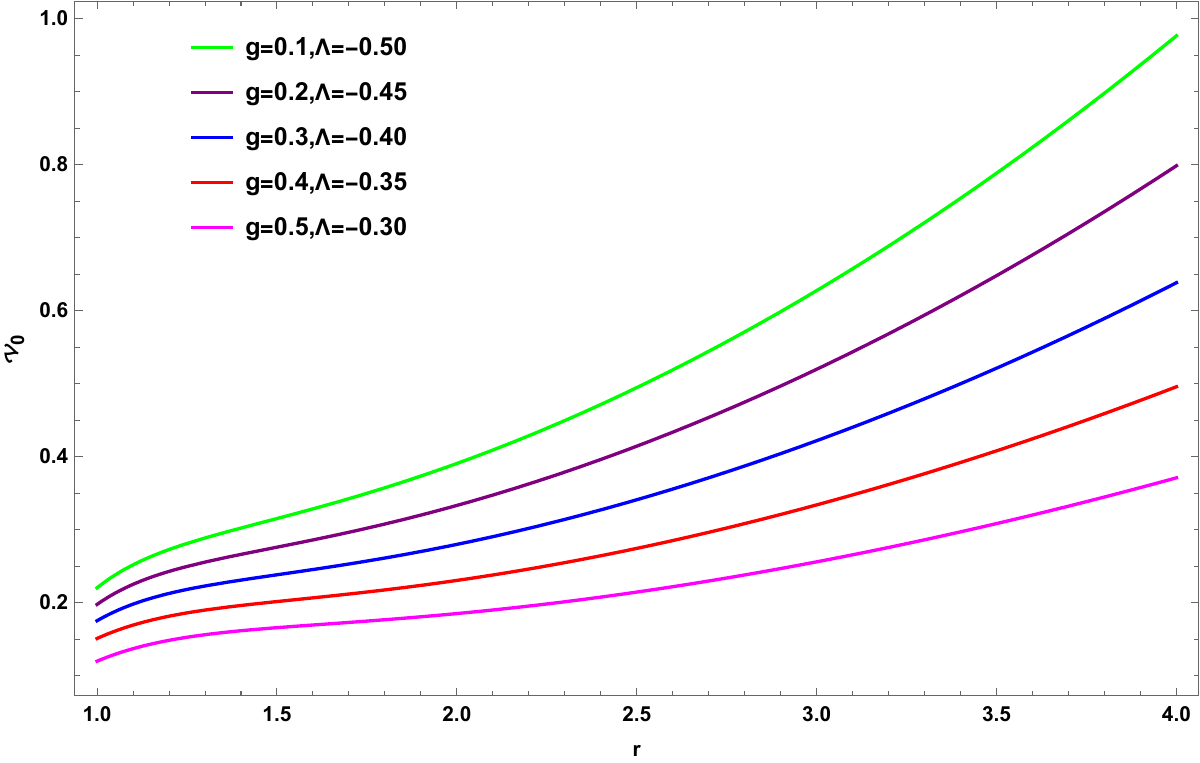}}\quad\quad
\subfloat[$g=0.1,\Lambda=-0.1,w=-2/3$]{\centering{}\includegraphics[scale=0.4]{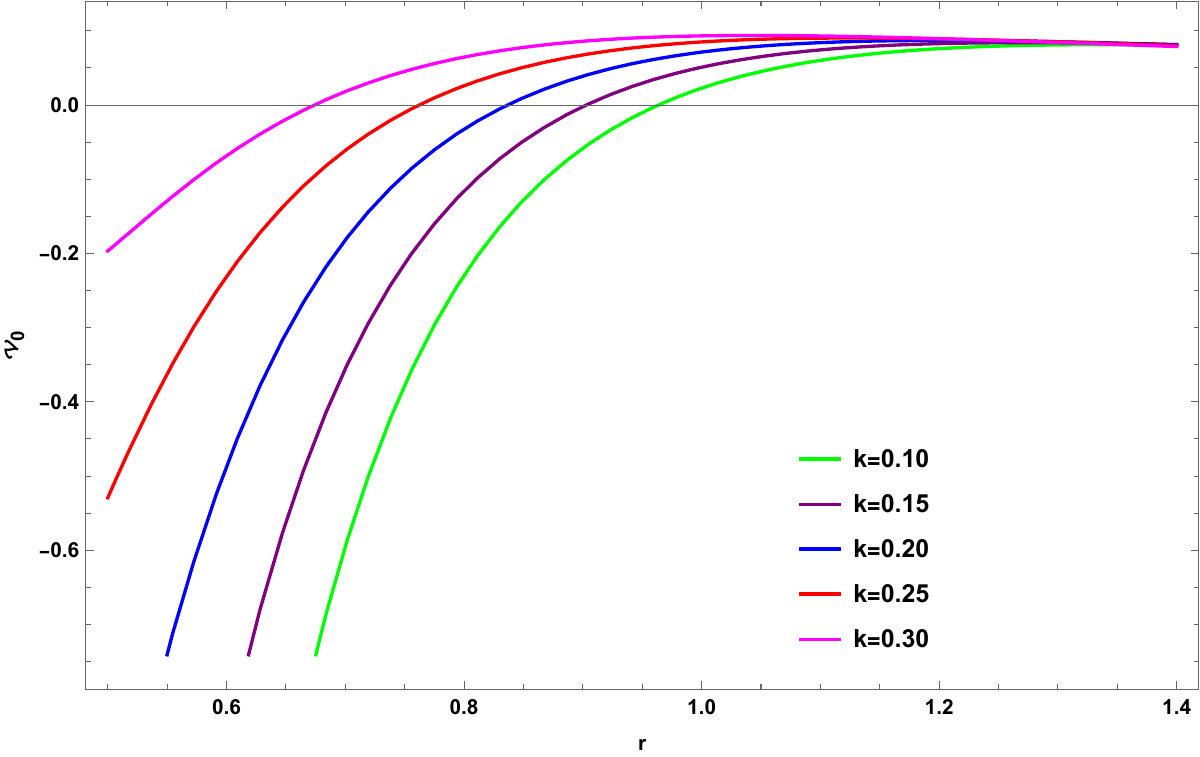}}
\hfill\\
\subfloat[$k=0.1=g,w=-2/3$]{\centering{}\includegraphics[scale=0.4]{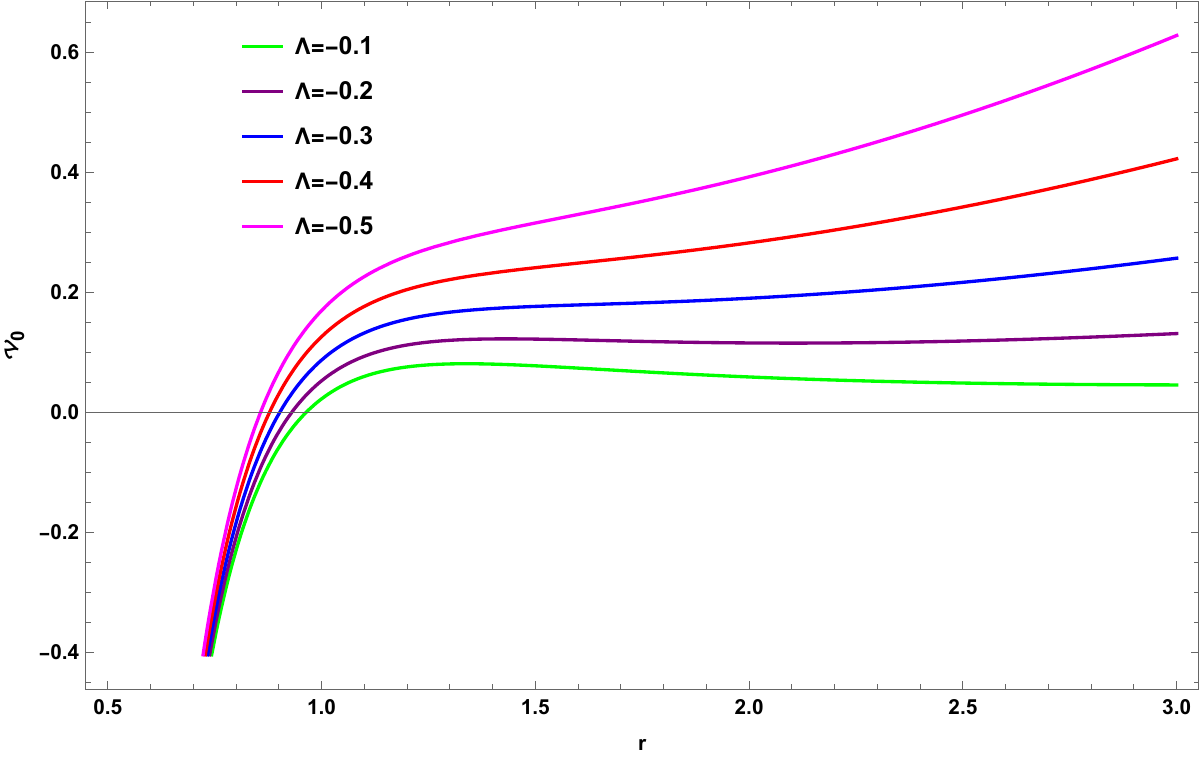}}\quad\quad
\subfloat[$\Lambda=-0.1,w=-2/3$]{\centering{}\includegraphics[scale=0.4]{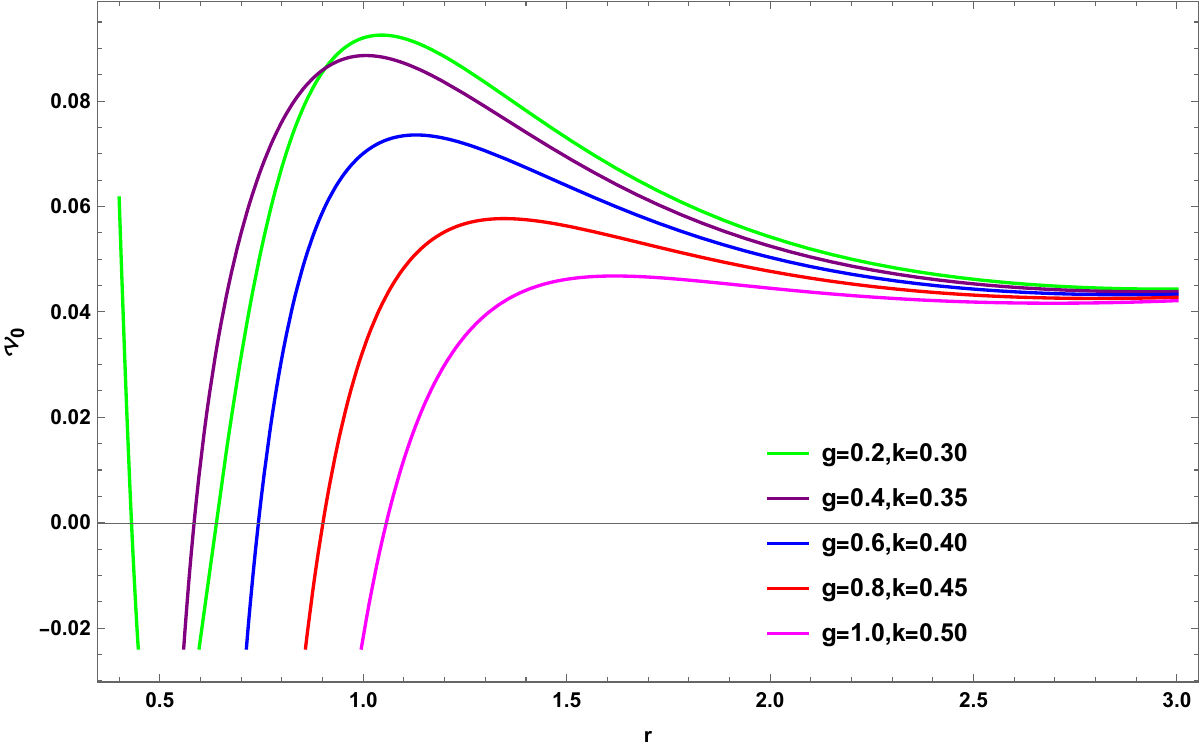}}
\hfill\\
\begin{centering}
\subfloat[$g=0.4,k=0.2,\Lambda=-0.1$]{\centering{}\includegraphics[scale=0.4]{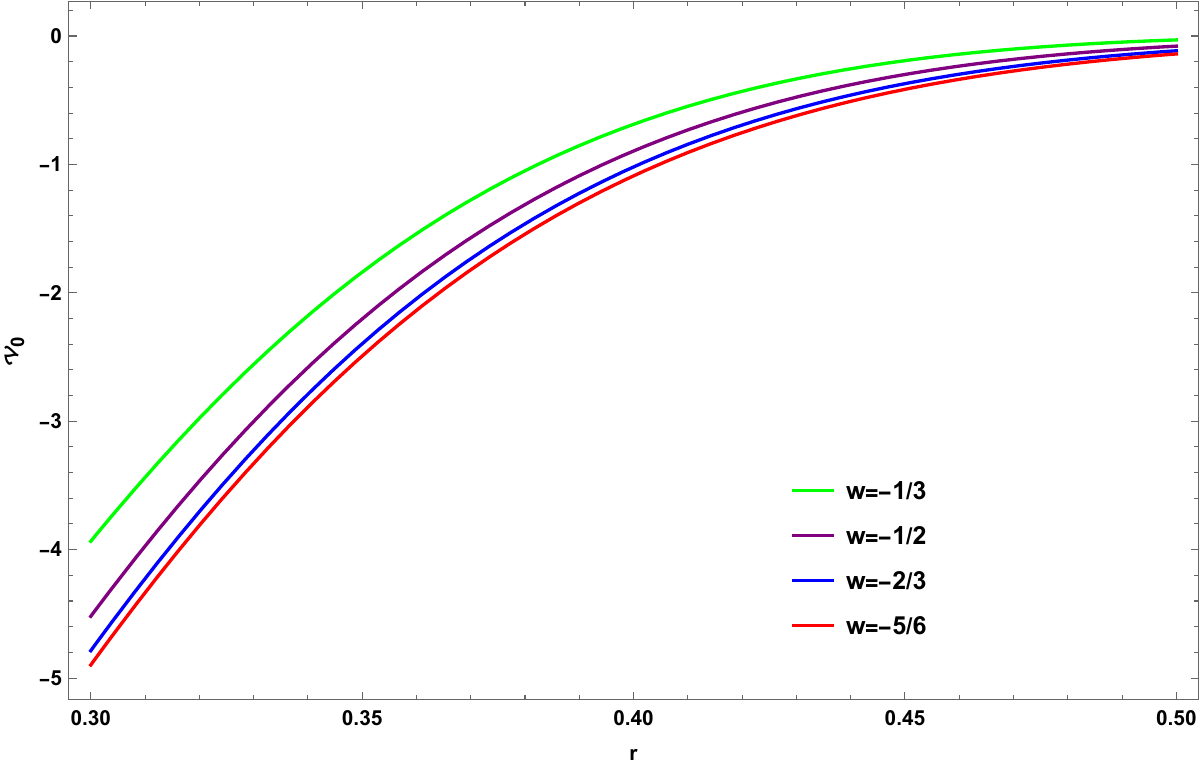}}
\par\end{centering}
\centering{}\caption{The RW-potential of Eq. (\ref{qq100}) for zero spin $s$-wave scalar field. Here we set $M=1/2, c=0.1$.}\label{fig:10}
\end{figure}
\par\end{center}

For an example, such as $s$-wave, where multipole number $\ell=0$, the RW-potential (\ref{qq10}) becomes
\begin{eqnarray}
    \mathcal{V}_0\Big{|}_{\ell=0}&=&\frac{1}{r}\,\left(1-\frac{2\,M\,r^2}{r^3+g^3}\,e^{-k/r}-\frac{\Lambda}{3}\,r^2-\frac{c}{r^{3\,w+1}}\right)\times\nonumber\\
    &&\Bigg[\frac{6\,e^{-k/r}\,M\,r^4}{(g^3 + r^3)^2}-\frac{2\,e^{-k/r}\,k\,M}{
       g^3 + r^3}-\frac{4\,e^{-k/r}\,M\,r}{g^3+r^3}- 
       c\,(-1-3\,w)\,r^{-2-3\,w}\Bigg].\label{qq100}
\end{eqnarray}

In Figure \ref{fig:10}, we have presented graphs that depict the behavior of the Regge-Wheeler (RW) potential given by equation (\ref{qq100}). These graphs illustrate the effects of various parameters, namely $k$, $g$, and $\mathrm{L}$ as well as the state parameter $w$, as functions of the radial coordinate $r$. The plots provide insight into how these parameters influence the RW-potential. 

\begin{center}
    {\bf Case B: One spin field: Electromagnetic Perturbations}
\end{center}

For spin-one, $S=1$ which corresponds to electromagnetic  perturbations, the RW-potential becomes
\begin{eqnarray}
    \mathcal{V}_1=\left(1-\frac{2\,M\,r^2}{r^3+g^3}\,e^{-k/r}-\frac{\Lambda}{3}\,r^2-\frac{c}{r^{3\,w+1}}\right)\,\frac{\ell\,(\ell+1)}{r^2}\label{qq11}
\end{eqnarray}
which reduced to the spin-one vector field potential for Hayward-Kiselev black hole under the case $k=0$ having state parameter $w$.

\begin{center}
    {\bf Case C: Spin-two field, $S=2$: Gravitational Perturbations}
\end{center}

For spin-two, $S=2$ which corresponds to gravitational  perturbations, the RW-potential becomes
\begin{eqnarray}
    \mathcal{V}_2&=&\left(1-\frac{2Mr^2}{r^3+g^3}e^{-k/r}-\frac{\Lambda}{3}r^2-\frac{c}{r^{3\,w+1}}\right)\Bigg[\frac{\ell(\ell+1)}{r^2}+\frac{2}{r^2}\Bigg\{-\frac{2Mr^2}{r^3+g^3}e^{-k/r}-\frac{\Lambda}{3}r^2-\frac{c}{r^{3\,w+1}}\Bigg\}\Bigg]\nonumber\\
    &-&\frac{1}{r}\,\left(1-\frac{2\,M\,r^2}{r^3+g^3}\,e^{-k/r}-\frac{\Lambda}{3}\,r^2-\frac{c}{r^{3\,w+1}}\right)\times\nonumber\\
    &&\Bigg(\frac{6\,e^{-k/r}\,M\,r^4}{(g^3 + r^3)^2}-\frac{2\,e^{-k/r}\,k\,M}{
       g^3 + r^3}-\frac{4\,e^{-k/r}\,M\,r}{g^3+r^3}- 
       c\,(-1-3\,w)\,r^{-2-3\,w}\Bigg).\label{qq12}
\end{eqnarray}

In our future attempt, we will calculate the quasinormal modes frequency by employing the First-Order WKB Approximation. This frequency expression is given by
\begin{equation}
    \omega^2 \approx \Bigg[\mathcal{V}_S(r_*)-i\,\left(n+\frac{1}{2}\right)\,\sqrt{-2\,\partial^2_{r_{*}}\,\mathcal{V}_S(r_*)} \Bigg]_{r_{*}=r_{*\,\,max}},\label{hh1}
\end{equation}
where $n \in \mathbf{N}$  is the overtone number, and  $r_{*}=r_{*\,\,max}$ is the tortoise coordinate location which maximise the relevant Regge-Wheeler potential. It is worth noting that $\mathcal{V}_S(r_*)\Big{|}_{r_{*}=r_{*\,\,max}}=\mathcal{V}_S(r)\Big{|}_{r=r_{\max}}$.

\section{Summary and Discussions}\label{sec:6}

NLEDs as a source of gravity, has the potential to produce regular black holes. In this paper, we present a new exact solution to the gravitational field equations in which NLEDs acts as  the source of matter in the presence of a QF. This solution describes a static, spherically symmetric black hole in the context of Anti-de Sitter (AdS) space-time, surrounded by quintessence matter. In other words, we obtained a solution which generalized the AdS-Schwarzschild-Kiselev black hole and, under certain limitations, reduces to the standard AdS-Schwarzschild as well as Bardeen-Kiselev black holes space-times. Our black hole solution can have at most two horizons: Cauchy and event horizons. As the value of quintessence parameter increase, so does the size of the black hole (event horizon). It does, however, diminish when the magnetic charge or deviation parameters grow. Furthermore, these parameters have critical values where there is no horizon radius for the critical values of $k$ and $g$ for a given state parameter $w$. 

We analyzed its thermodynamic properties numerically and graphically. The temperature increases in proportion to the deviation and quintessence parameters. We have seen that the temperature of this regular black hole rises sharply to a maximum value for a certain horizon radius, then it falls exponentially as the horizon radius increases, and finally rises as the horizon radius approaches infinity. We also investigated thermodynamic stability by examining heat capacity behavior. Our thorough examination of the specific heat functions revealed that the black hole can exist in three phases: stable, unstable, and stable. We then focused on geodesics motions of test particles around this regular black hole solution, demonstrating how the NLEDs and QF including the angular momentum affects the effective potential $V_{eff}$ for both massless and massive particles for a given state parameter $w$ in Figures \ref{fig:5} to \ref{fig:6}. The effective potential of this black hole is shown to rise when the magnetic charge or deviation parameters increase. However, it decreases as the state parameter of QF increases. Moreover, we obtained an expression of force on the massless photon particle and demonstrated that both NLED and QF parameters represented by $k,g$ including the conserved angular momentum $\mathrm{L}$ influences on it for a particular state parameter $w$. Graphically in Figure (\ref{fig:8}), we illustrated this force for different values of $k,g,\mathrm{L}$ as well as the state parameter $w$.  Finally, we investigated the RW potential for this regular black hole and showed how modifications in the parameters (NLEDs and QF) affect the RW-potential for all fields with different spins. 

%In summary, we found that the parameters of NLED sources and QF in modified gravity affect the thermodynamic properties, geodesics and RW potential of this regular black hole. Furthermore, it would be interesting to examine the properties of the NLEDs black holes with QF coupled with their sources of different modified gravity.  

In Ref. \cite{LCNS}, the authors obtained new exact solutions for the gravitational field equations within the framework of $f(R,T)$ gravity. These solutions describe various classes of black holes surrounded by fluids, based on specific values of the equation of state parameter $w$ (which takes values such as 0, -1/3, -2/3, -1, and -4/3). Notably, this non-rotating black hole solution is the modifications of the Kiselev black hole within the context of $f(R,T)$ gravity and serve to reproduce well-known solutions of the Einstein field equations as exceptional cases. Building on this work, the authors in Ref. \cite{SGG} derived a rotating black hole solution in $f(R,T)$ gravity from the spherical Kiselev black hole, employing the revised Newman-Janis algorithm. The resulting black hole metric, known as the $f(R,T)$-motivated rotating Kiselev black hole (FRKBH) metric, reduces to the Kerr black hole under the special case $K=0$, as discussed in their study. The authors also analyzed specific values of the state parameter $w$ and presented their results. In this article, we examined a non-rotating black hole coupled with nonlinear electrodynamics and surrounded by a quintessence field. In future work, we plan to extend this study to rotating black holes in modified gravity theories, such as $f(R,T)$ and $f(R,G,T)$, employing robust methods like the Newman-Janis algorithm and analyzing the resulting solution.

\small

\section*{Data Availability Statement}

No new data were generated or analyzed in this study.

\section*{Conflicts of Interest}

There is no conflict of interests.

\section*{Funding Statement}

No fund has received for this paper.

\section*{Code/Software}

No software/Coder were used in this study.

\section*{Acknowledgments}

F.A. acknowledges the Inter University Centre for Astronomy and Astrophysics (IUCAA), Pune, India for granting visiting associateship.

\normalsize

\end{document}